\documentclass[prd, twocolumn, superscriptaddress]{revtex4-2}
\usepackage{amsmath}
\usepackage{amssymb}
\usepackage{dsfont}
\usepackage{graphicx}
\usepackage{xcolor}
\usepackage{tikz}
\usetikzlibrary{positioning}
\allowdisplaybreaks[1]

\begin{document}
\title{Minkowski-Fock states in accelerated frames}

\author{Riccardo Falcone}
\affiliation{Department of Physics, University of Sapienza, Piazzale Aldo Moro 5, 00185 Rome, Italy}

\author{Claudio Conti}
\affiliation{Department of Physics, University of Sapienza, Piazzale Aldo Moro 5, 00185 Rome, Italy}
\affiliation{Institute for Complex Systems (ISC-CNR), Department of Physics, University Sapienza, Piazzale Aldo Moro 2, 00185, Rome, Italy}
\affiliation{Research Center Enrico Fermi, Via Panisperna 89a, 00184 Rome, Italy}

\begin{abstract}
An explicit Wigner formulation of Minkowski particle states for non-inertial observers is unknown. Here, we derive a general prescription to compute the characteristic function for Minkowski-Fock states in accelerated frames. For the special case of single-particle and two-particle states, this method enables to derive mean values of particle numbers and correlation function in the momentum space, and the way they are affected by the acceleration of the observer. We show an indistinguishability between Minkowski single-particle and two-particle states in terms of Rindler particle distribution that can be regarded as a way for the observer to detect any acceleration of the frame. We find that for two-particle states the observer is also able to detect acceleration by measuring the correlation between Rindler particles with different momenta.
\end{abstract}

\maketitle

\section{Introduction}

We investigate the general expression of the characteristic function \cite{PhysRev.177.1857,PhysRev.177.1882} for particle states emitted in a Minkowski space-time and registered by an accelerated observer. The aim is to provide a comprehensive description for Minkowski-Fock states in non-inertial frames \cite{Barnett2002}.

In a recent paper \cite{2020}, Ben-Benjamin, Scully, and Unruh reported the Wigner distribution for Minkowski-Fock states in the right and left Rindler wedges. However, to the best of our knowledge, the case of the right Rindler wedge -- i.e. as the result of the partial trace over the left wedge -- is still missing.

The characteristic function of any state can be used to obtain expectation values through simple derivatives. However, computing the characteristic function in the Rindler space-time requires a series of non-trivial theoretical properties arising from the transformation of the state from the inertial to the accelerated frame. Here we show that these rules can be formulated in a way such that one can build algorithmically a general expression of states with arbitrary number of particles. We also give a diagrammatic representation of the characteristic function resulting from our combinatorial method.

As an application of the general methodology, we consider single-particle and two-particle states and extract the probability distribution to find a Rindler particle with a specific momentum and the correlation between Rindler particles with different momenta. We also provide some example with Gaussian-like wave-functions.

The probability distribution and the correlation function changes from an inertial to an accelerated frame. For specific choices of wave-functions, the single-particle probability distribution becomes indistinguishable from the two-particle case, at variance with what happens in the Minkowski space-time. While in the case of single-particle states the correlation between different momenta has the same form in both the Minkowski and the Rindler space-time -- with the exception of the vacuum correlations -- in the case of two-particle states they differ. This suggests a way to distinguish between single-particle and two-particle states for non-inertial observers. This also implies that one can exploit Minkowski two-particle states to measure the acceleration of the observer.

The paper is organized as follows. In Sec. \ref{Result} we show the general expression of the characteristic function for general Minkowski-Fock states. In Sec. \ref{Method} we briefly describe the method to obtain the characteristic function. Both the solutions and the method are provided in the 1+1 dimensional case, while arguments for the possibility to extend the same results in 3+1 dimensions are given in Sec. \ref{1_3_massive_field}. Sections \ref{Single-particle state}, \ref{Single-particle_state_in_3+1_dimensions} and \ref{Two-particle state} report the application to the case of single and two-particle states, including the explicit characteristic function, the observable quantities and the example of Gaussian wave-functions. Conclusions are drawn in Sec. \ref{Conclusions}. Full details for the analytical results are given in Appendix \ref{A_proof_for} and \ref{A_proof_for_2}. 

\section{General expression of the characteristic function}\label{Result}

By following Fulling, Davis and Unruh \cite{PhysRevD.7.2850, 1975, Unruh}, we consider a (1+1)-dimensional flat space-time $(t,x)$ and a massless free scalar field $\hat{\phi}(t,x)$. The accelerated frame with acceleration $a c^2>0$ -- where $c$ is the speed of light -- can be described by a coordinate patch $(T_R,X_R)$ with the following coordinate transformation: $a c t = \exp (aX_R) \sinh (a c T_R)$ and $ a x = \exp (aX_R) \cosh (a c T_R) $. Such transformation covers the right Rindler wedge, defined by $x > c |t|$. The left Rindler wedge (defined by $x < - c |t|$) can be covered by the transformation $a c t = \exp(-aX_L) \sinh (a c T_L)$ and $ a x = - \exp(-aX_L) \cosh (a c T_L) $. We define $\hat{a}(k)$ as the annihilation operator for the Minkowski mode with momentum $k$, and $\hat{b}_L(K)$ ($\hat{b}_R(K)$) as the annihilation operators for the left (right) Rindler mode with momentum $K$.

Minkowski-Fock pure states $|\psi\rangle$ are elements of the $\hat{a}(k)$-algebra representation space and they can also be represented as elements of the $\hat{b}_R(K)$-algebra representation space through mixed states $\hat{\rho}$ of the right-Rindler-Fock space by using the following procedure. We write any $|\psi\rangle$ as a combination of chains of creator operators $\hat{a}^\dagger(k)$ acting on the Minkowski vacuum state $|0_M\rangle $:
\begin{equation}\label{characteristic_function_step_1}
|\psi\rangle = \sum_{n} \psi[n]  |n \rangle,
\end{equation}
where $\psi[n]$ is the probability amplitude to find $|\psi\rangle $ in the following non-normalized Minkowski-Fock state
\begin{equation}\label{characteristic_function_step_1_2}
|n \rangle = \prod_k \left[ \hat{a}^\dagger(k) \right]^{n(k)} |0_M \rangle
\end{equation}
and the sum in Eq. (\ref{characteristic_function_step_1}) can be identified with a generalized sum -- i.e. both discrete sums and integrals -- over all Minkowski-Fock states. $\hat{a}^\dagger(k)$-operators can be converted in Rindler creation $\hat{b}_{L,R}^\dagger(K)$ and annihilation $\hat{b}_{L,R}(K)$ operators thanks to the following Bogolyubov transformation \cite{Mukhanov2007}:
\begin{align}\label{Bogolyubov_transformation}
\hat{a}(k) = & \int_{-\infty}^{+\infty}dK \left[ \alpha(k,K)\hat{b}_L(K) - \beta^*(k,K) \hat{b}^\dagger_L(K) \right. \nonumber \\
& \left. + \alpha^*(k,K) \hat{b}_R(K) - \beta(k,K) \hat{b}^\dagger_R(K) \right],
\end{align}
with
\begin{subequations}\label{Bogolyubov_coefficients}
\begin{equation}
\alpha(k,K) = \theta(kK) \sqrt{\frac{K}{k}}F(k,K),
\end{equation}
\begin{equation}
\beta(k,K) = \theta(kK) \sqrt{\frac{K}{k}}F(-k,K),
\end{equation}
\begin{align} \label{F}
F(k,K) = & \frac{1}{2 \pi a} \Gamma \left( -\frac{i K}{a} \right) \nonumber \\
& \times \exp \left( i \frac{K}{a} \ln \frac{|k|}{a} + \text{sign} \left( k \right) \frac{\beta}{4} K \right)
\end{align}
\end{subequations}
and $\beta = 2 \pi/a$. Moreover, $|0_M\rangle $ can be written as an element of the Rindler-Fock space -- i.e. $\hat{b}_{L,R}(K)$-algebra representation space -- thanks to the following identity
\begin{align}  \label{vacuum_state}
|0_M\rangle \propto & \exp \left( \int_{-\infty}^{+\infty} dK \exp \left( - \frac{\beta}{2} |K| \right) \right. \nonumber \\
& \times \left. \hat{b}^\dagger_L(K) \hat{b}^\dagger_R(K) \right) | 0_L, 0_R \rangle,
\end{align}
which in turns is the result of the definition $\hat{a}(k) |0_M\rangle = 0$ and the Bogolyubov transformation of Eq. (\ref{Bogolyubov_transformation}). Eqs. (\ref{characteristic_function_step_1}, \ref{Bogolyubov_transformation}, \ref{vacuum_state}) allow to represent $|\psi\rangle$ as an element of the Rindler-Fock space. Finally the partial trace over the left wedge can be performed in order to obtain $\hat{\rho}$:
\begin{equation}
\hat{\rho} = \text{Tr}_L |\psi\rangle \langle \psi |.
\end{equation}

In the specific case of Minkowski vacuum state $| \psi \rangle = | 0_M \rangle$, the statistical operator $\hat{\rho}$ is identified by the following thermal state \cite{Unruh}
\begin{equation}\label{rho_0}
\hat{\rho}_0 \propto \sum_{n} \left[ \int_{-\infty}^{+\infty} dK e^{-\beta |K|} n(K) \right] | n \rangle \langle n|,
\end{equation}
where, in this case,
\begin{equation}\label{n_rho_0}
|n \rangle = \prod_K \left[ \hat{b}_R^\dagger(K) \right]^{n(K)} |0_L, 0_R \rangle.
\end{equation}

An equivalent representation for Minkowski-Fock states in the accelerated frame can be made through the following definition of characteristic function in the right-Rindler space-time \cite{Barnett2002}:
\begin{equation} \label{chi}
\chi^{(p)}[\xi, \xi^*] = \text{Tr} \left( \hat{\rho} \hat{D}_p[\xi, \xi^*] \right),
\end{equation}
where $\xi = \xi(K)$ is a complex function, $p$ can take values $-1$, $0$ and $+1$ and
\begin{align}
 \hat{D}_p[\xi, \xi^*] = & \exp \left( \int_{-\infty}^{+\infty} dK \left[ \xi(K) \hat{b}^\dagger_R(K) \right.  \right. \nonumber \\
 & \left.\left.  - \xi^*(K) \hat{b}_R(K) + \frac{p}{2} |\xi(K)|^2\right] \right).
\end{align}
Specifically for $p=-1$
\begin{align}\label{D_-1}
\hat{D}_{-1}[\xi, \xi^*] = & \exp \left(- \int_{-\infty}^{+\infty} dK \xi^*(K) \hat{b}_R(K) \right) \nonumber \\
& \times \exp \left( \int_{-\infty}^{+\infty} dK  \xi(K) \hat{b}^\dagger_R(K)\right).
\end{align}

In the present paper we show how to obtain an explicit expression for $\chi^{(p)}[\xi, \xi^*]$. The result is the following:\begin{subequations}\label{chi_final}
\begin{equation}\label{chi_final_a}
\chi^{(p)}[\xi, \xi^*] = \sum_{n, n'} \psi[n] \psi^*[n'] \bar{\chi}^{(p)} (\mathbf{K}(n), \mathbf{K}(n')) [\xi, \xi^*],
\end{equation}
\begin{align} \label{chi_bar_final}
 & \bar{\chi}^{(p)}(\mathbf{U}, \mathbf{U}')[\xi, \xi^*] = \chi^{(p)}_0[\xi,\xi^*] \sum_{\substack{\mathbf{S} \subseteq \mathbf{U} \\ \mathbf{S}' \subseteq \mathbf{U}' }} c(\mathbf{U} \setminus \mathbf{S},\mathbf{U}' \setminus \mathbf{S}')  \nonumber \\
 & \times \prod_{k\in\mathbf{S}}  \{ - L (k) [\xi,\xi^*]\}^*  \prod_{k'\in\mathbf{S}'}L (k') [\xi,\xi^*]  ,
\end{align}
\end{subequations}
where $\mathbf{K}(n)$ is the space of momenta $k$ repeated $n(k)$ times and the sum of Eq. (\ref{chi_bar_final}) runs over all the possible subsets $\mathbf{S}\subseteq \mathbf{U}$ and $\mathbf{S}'\subseteq \mathbf{U}'$. The coefficients $c(\mathbf{U} \setminus \mathbf{S},\mathbf{U}' \setminus \mathbf{S}') $ of Eq.(\ref{chi_final}) have the following combinatoric expression
\begin{equation} \label{c_combinatory}
c \left( \{k_i\}_{i=1}^M,\{k'_{j}\}_{j=1}^N \right) = \delta_{MN} \sum_\mathcal{P} \prod_{i} \delta(k_i - k'_{\mathcal{P}(i)}),
\end{equation}
where $\{k_i\}_{i=1}^M$ and $\{k'_{j}\}_{j=1}^N$ are any arbitrarily ordered sequence of elements in $\mathbf{U} \setminus \mathbf{S}$ and $\mathbf{U}' \setminus \mathbf{S}'$ and the sum runs over all the possible permutations $\mathcal{P}$ for the index $i$. Finally, $L (k) [\xi,\xi^*]  $ of Eq.(\ref{chi_bar_final}) is a linear functional of $\xi$ and $\xi^*$ defined as
\begin{equation}
L (k) [\xi,\xi^*] = \int_{-\infty}^{+\infty} dK \left[\alpha^*(k,K) \xi(K) - \beta(k,K) \xi^*(K) \right].
\end{equation}
It can be noticed that $L (k) [\xi,\xi^*]$ appears also in Eq. (\ref{Bogolyubov_transformation}) as a Bogolyubov transformation between $\hat{b}_{L,R}(K)$- and $\hat{a}(k)$-operators:
\begin{equation}
\hat{a}(k) = \left\lbrace L (k) \left[\hat{b}^\dagger_L(K),\hat{b}_L(K)\right] \right\rbrace^\dagger + L (k) \left[\hat{b}_R(K),\hat{b}^\dagger_R(K)\right]
\end{equation}

The explicit form of $\chi^{(p)}[\xi, \xi^*]$ given by Eq. (\ref{chi_final}) can be compared with the explicit form of the characteristic function of $|\psi\rangle \langle \psi |$ in the Minkowski space-time, which, in turn, is defined in the following way
\begin{align} \label{chi_M}
\chi^{(p)}_M[\xi_M, \xi^*_M] = & \text{Tr} \left( |\psi\rangle \langle \psi | \exp \left( \int_{-\infty}^{+\infty} dk \left[ \xi_M(k) \hat{a}^\dagger(k) \right.  \right.  \right. \nonumber \\
 &  \left.\left.\left.  - \xi_M^*(k) \hat{a}(k) + \frac{p}{2} |\xi_M(k)|^2\right] \right) \right).
\end{align}
An explicit form for $\chi^{(p)}_M[\xi_M, \xi^*_M]$ of Eq. (\ref{chi_M}) has been computed in Appendix \ref{A_proof_for_2} and reads
\begin{widetext}
\begin{equation}\label{chi_M_final}
\chi^{(p)}_M[\xi_M, \xi_M^*] = \chi^{(p)}_{0M}[\xi_M,\xi_M^*] \sum_{n, n'} \psi[n] \psi^*[n'] \sum_{\substack{\mathbf{S} \subseteq \mathbf{K}(n) \\ \mathbf{S}' \subseteq \mathbf{K}(n') }} c(\mathbf{K}(n) \setminus \mathbf{S},\mathbf{K}(n') \setminus \mathbf{S}')  \prod_{k\in\mathbf{S}} \left[ - \xi^*_M(k) \right]  \prod_{k'\in\mathbf{S}'}  \xi_M(k'),
\end{equation}
\end{widetext}
where $\chi^{(p)}_{0M}[\xi_M,\xi_M^*]$ is the characteristic function of $|0_M \rangle \langle 0_M |$ in the Minkowski space-time
\begin{align} \label{chi_0M}
\chi^{(p)}_{0M}[\xi_M, \xi^*_M] = & \text{Tr} \left( |0_M\rangle \langle 0_M | \exp \left( \int_{-\infty}^{+\infty} dk \left[ \xi_M(k) \hat{a}^\dagger(k) \right.  \right.  \right. \nonumber \\
 &  \left.\left.\left.  - \xi_M^*(k) \hat{a}(k) + \frac{p}{2} |\xi_M(k)|^2\right] \right) \right),
\end{align}
with the following explicit form
\begin{equation} \label{chi_0M_final}
\chi^{(p)}_{0M}[\xi_M, \xi^*_M] = \exp \left( \int_{-\infty}^{+\infty} dk \frac{p-1}{2} |\xi_M(k)|^2 \right).
\end{equation}
By comparing Eq. (\ref{chi_M_final}) with Eq. (\ref{chi_final}) it is possible to see that the transformation of the characteristic function from the Minkowski to the right-Rindler space-time $\chi^{(p)}_M[\xi_M, \xi_M^*] \mapsto \chi^{(p)}[\xi, \xi^*]$ can be easily computed by performing the following substitutions in Eq. (\ref{chi_M_final}): $ \chi^{(p)}_{0M}[\xi_M,\xi_M^*] \mapsto \chi^{(p)}_0[\xi,\xi^*]$ and $\xi_M(k) \mapsto  L (k) [\xi,\xi^*] $.

Finally we want to show that Eq. (\ref{chi_bar_final}) can be put in a diagrammatic form by defining a single diagram through the following procedure. Write all the elements of $\mathbf{U}$ and $\mathbf{U}'$ in two distinct columns and create some pair connections between elements of the left and the right column $k_i \text{ --- } k'_j$. The numerical value associated to this diagram is the product of $\chi^{(p)}_0[\xi,\xi^*]$ and the following contributions coming from the elements of the diagram. Each pair connection $k_i \text{ --- } k'_j$ contributes with a delta function between the two momenta $\delta(k_i-k'_j)$. Each left-column element $k_i$ left without pair connection contributes with $\{ - L (k_i) [\xi,\xi^*] \}^*$. On the other hand, a ``free'' right-column element $k'_i$ contributes with $ L (k'_i) [\xi,\xi^*]$. In this way, a diagrammatic expression of Eq. (\ref{chi_bar_final}) is the following:
\begin{widetext}
\begin{equation}
\begin{tikzpicture}
\node (0_1) [] {$ \bar{\chi}^{(p)}(\{k_i\}_{i=1}^M,\{k'_{j}\}_{j=1}^N)[\xi, \xi^*] = $};

\node (1L_vdots) [right=0cm of 0_1] {$\vdots$};
\node (1R_vdots) [right of=1L_vdots] {$\vdots$};

\node (1L_k_1) [above=0cm of 1L_vdots] {$k_1$};
\node (1L_k_M) [below=0cm of 1L_vdots] {$k_M$};
\node (1R_k_1) [above=0cm of 1R_vdots] {$k'_1$};
\node (1R_k_N) [below=0cm of 1R_vdots] {$k'_N$};

\node (1_2) [right of=1R_vdots] {$+ \sum_{ii'}$};
\node (2L_k_i) [right of=1_2] {$k_i$};
\node (2R_k_i) [right of=2L_k_i] {$k'_{i'}$};

\node (2L_vdots_1) [above=0cm of 2L_k_i] {$\vdots$};
\node (2L_k_1) [above=0cm of 2L_vdots_1] {$k_1$};
\node (2L_vdots_2) [below=0cm of 2L_k_i] {$\vdots$};
\node (2L_k_M) [below=0cm of 2L_vdots_2] {$k_M$};
\node (2R_vdots_1) [above=0cm of 2R_k_i] {$\vdots$};
\node (2R_k_1) [above=0cm of 2R_vdots_1] {$k'_1$};
\node (2R_vdots_2) [below=0cm of 2R_k_i] {$\vdots$};
\node (2R_k_N) [below=0cm of 2R_vdots_2] {$k'_N$};

\node (2_3) [right of=2R_k_i] {$+ \sum_{ii'jj'}$};
\node (3L_vdots) [right of=2_3] {$\vdots$};
\node (3R_vdots) [right of=3L_vdots] {$\vdots$};

\node (3L_k_i) [above=0cm of 3L_vdots] {$k_i$};
\node (3L_vdots_1) [above=0cm of 3L_k_i] {$\vdots$};
\node (3L_k_1) [above=0cm of 3L_vdots_1] {$k_1$};
\node (3L_k_j) [below=0cm of 3L_vdots] {$k_j$};
\node (3L_vdots_2) [below=0cm of 3L_k_j] {$\vdots$};
\node (3L_k_M) [below=0cm of 3L_vdots_2] {$k_M$};
\node (3R_k_i) [above=0cm of 3R_vdots] {$k'_{i'}$};
\node (3R_vdots_1) [above=0cm of 3R_k_i] {$\vdots$};
\node (3R_k_1) [above=0cm of 3R_vdots_1] {$k'_1$};
\node (3R_k_j) [below=0cm of 3R_vdots] {$k'_{j'}$};
\node (3R_vdots_2) [below=0cm of 3R_k_j] {$\vdots$};
\node (3R_k_N) [below=0cm of 3R_vdots_2] {$k'_N$};

\node (3_4) [right of=3R_vdots] {$+ \dots$};

\draw[-] (2L_k_i) to (2R_k_i);
\draw[-] (3L_k_i) to (3R_k_i);
\draw[-] (3L_k_j) to (3R_k_j);
\end{tikzpicture}
\end{equation}
\end{widetext}

\section{Method}\label{Method}

In this section we show how to obtain Eqs. (\ref{chi_final}), given some properties of the Minkowski vacuum state $|0_M\rangle$ and its representation in the right-Rindler wedge $\hat{\rho}_0$, which has the explicit form given by Eqs. (\ref{rho_0}, \ref{n_rho_0}). The properties of $|0_M\rangle$ and $\hat{\rho}_0 $ that we exploit are the following:
\begin{itemize}

\item  the creation of a Rindler particle in the left (right) wedge over the Minkowski vacuum background is equivalent to the destruction of a Rindler particle in the right (left) wedge, up to an $\exp(\beta|K|/2)$ factor: \begin{equation} \label{vacuum_state_property}
\hat{b}^\dagger_{L,R}(K) |0_M\rangle = \exp \left( \frac{\beta}{2} |K| \right) \hat{b}_{R,L}(K) |0_M\rangle
\end{equation}
(see Appendix \ref{A_proof_for} for the proof);

\item it is possible to move creation $\hat{b}_R^\dagger(K)$ and annihilation $\hat{b}_R(K)$ operators acting from the left of $\hat{\rho}_0$ to its right and the other way round by using the following identity and its adjoint
\begin{equation}\label{ordering_Rindler}
\hat{b}^\dagger_R(K) \hat{\rho}_0 = e^{\beta |K|}  \hat{\rho}_0 \hat{b}^\dagger_R(K)
\end{equation}
(see Appendix \ref{A_proof_for} for the proof);

\item the functional derivatives of
\begin{equation}
\chi^{(p)}_0[\xi,\xi^*] = \text{Tr} \left( \hat{\rho}_0 \hat{D}_p[\xi, \xi^*] \right)
\end{equation}
for $p=-1$ with respect to different $\xi(K)$ and $\xi^*(K)$ give the following mean values
\begin{align}\label{multiple_Xi_derivatives_-1}
& \prod_K \left[ \frac{\delta}{\delta \xi(K)} \right]^{M(K)} \prod_{K'} \left[- \frac{\delta}{\delta \xi^*(K')} \right]^{N(K')}  \chi^{(-1)}_0 [\xi,\xi^*] \nonumber \\
= & \text{Tr} \left( \hat{\rho}_0 \prod_{K'} \left[\hat{b}(K')\right]^{N(K')} \hat{D}_{-1} [\xi,\xi^*] \prod_K \left[\hat{b}^\dagger(K)\right]^{M(K)} \right)
\end{align}
for any $M(K)$ and $N(K')$, as it can be noticed from Eq. (\ref{D_-1});

\item $\chi^{(p)}_0[\xi,\xi^*]$ is already known in literature \cite{Barnett2002}, since $\hat{\rho}_0$ has the form of a thermal state
\begin{align} \label{thermal_characteristic_fucntion}
\chi^{(p)}_0[\xi,\xi^*] = & \exp \left( \int_{-\infty}^{+\infty}dK |\xi(K)|^2 \right. \nonumber \\
& \left. \times \left[- n_0(K) + \frac{p-1}{2} \right] \right),
\end{align}
with $n_0(K) = (e^{\beta|K|}-1)^{-1}$.

\end{itemize}

Given such information, it is possible to provide a generic procedure in order to put $\chi^{(p)}[\xi,\xi^*]$ in the form of Eqs. (\ref{chi_final}). Such procedure follows the following steps:
\begin{enumerate}

\item express the $\hat{a}(k)$-operators of Eq. (\ref{characteristic_function_step_1_2}) in terms of $\hat{b}_{L,R}(K)$-operators using the transformation (\ref{Bogolyubov_transformation}), so that $|\psi\rangle \langle \psi |$ is put in the form of a combination of chains of $\hat{b}_{L,R}(K)$-operators acting from the left and right of $|0_M \rangle \langle 0_M |$; \label{prescription_2}

\item convert all $\hat{b}_L(K)$-operators acting on $|0_M \rangle \langle 0_M |$ into $\hat{b}_R(K)$-operators using Eq. (\ref{vacuum_state_property}), so that $|\psi\rangle \langle \psi |$ is put in the form of a combination of chains of $\hat{b}_R(K)$-operators acting from the left and right of $|0_M \rangle \langle 0_M |$; \label{prescription_3}

\item perform the partial trace over the left wedge, so that $\hat{\rho}$ is put in the form of a combination of chains of $\hat{b}_R(K)$-operators acting from the left and right of $\hat{\rho}_0$; \label{prescription_4}

\item reorder the $\hat{b}_R(K)$-operators using Eq. (\ref{ordering_Rindler}) and the canonical commutating rules by choosing the rearrangement such that $\hat{\rho}$ is put in the form of a combination of chains of creation operators $ \hat{b}_R^\dagger(K)$ acting from the left of $\hat{\rho}_0$ and annihilation operators $ \hat{b}_R(K)$ from the right of $\hat{\rho}_0$; \label{prescription_5}

\item multiply such linear combination with $\hat{D}_{-1}[\xi,\xi^*]$, use the trace over the right wedge and the cyclic property of the trace in order to see $\chi^{(-1)}[\xi,\xi^*]$ as a combination of terms that have the same form of the right side of Eq. (\ref{multiple_Xi_derivatives_-1}); \label{prescription_6}

\item by using Eq. (\ref{multiple_Xi_derivatives_-1}), read $\chi^{(- 1)}[\xi,\xi^*]$ in terms of linear combinations of multiple $\xi$-derivatives of $\chi^{(- 1)}_0[\xi,\xi^*]$, which are explicitly obtainable from Eq. (\ref{thermal_characteristic_fucntion}); \label{prescription_7}

\item extract the final expression of $\chi^{(p)}[\xi,\xi^*]$ from $\chi^{(- 1)}[\xi,\xi^*]$ through the following multiplication
\begin{equation}
\chi^{(p)}[\xi,\xi^*] = \exp \left( \int_{-\infty}^{+\infty} dK \frac{p+1}{2} |\xi(K)|^2 \right) \chi^{(- 1)}[\xi,\xi^*],
\end{equation}
which can be easily performed by replacing $\chi^{(- 1)}_0[\xi,\xi^*]$ of step \ref{prescription_7} with $\chi^{(p)}_0[\xi,\xi^*]$. \label{prescription_8}

\end{enumerate}
In Appendix \ref{A_proof_for} we give the full details for the proof which leads to Eqs. (\ref{chi_final}).

\section{3+1 dimensions} \label{1_3_massive_field}

Only in the case of 1+1 dimensions, the scalar field is free in the Rindler frames. Indeed, the massless Klein-Gordon equation in a conformally flat metric is equivalent to the flat space-time case only for 1+1 space-time dimensions. For such reason, in the previous sections, we have considered Rindler particles with energy $|K|$ having a defined momentum $K$. This does not happen in 3+1 dimensions, where Rindler particles with energy $\Omega>0$ can only have defined momentum components along the direction orthogonal to the acceleration. In the 3+1 dimensional case we are, therefore, forced to identify Rindler particles with the energy $\Omega$ and the transverse momentum $\textbf{K}_\perp$ along the $y$ and $z$ axis.

However, in the 3+1 dimensional case, our method remains the same, with the difference that Eq. (\ref{vacuum_state_property}) is replaced by \cite{RevModPhys.80.787}
\begin{equation}\label{vacuum_state_property_1_3}
\hat{b}^\dagger_{L,R}(\Omega,\textbf{K}_\perp) |0_M\rangle = \exp \left( \frac{\beta}{2} \Omega \right) \hat{b}_{R,L}(\Omega, -\textbf{K}_\perp) |0_M\rangle,
\end{equation}
Eq. (\ref{ordering_Rindler}) by
\begin{equation}\label{ordering_Rindler_1_3}
\hat{b}^\dagger_R(\Omega,\textbf{K}_\perp) \hat{\rho}_0 = e^{\beta \Omega}  \hat{\rho}_0 \hat{b}^\dagger_R(\Omega,\textbf{K}_\perp),
\end{equation}
and the new Bogolyubov coefficients $\alpha(\vec{k},\Omega,\textbf{K}_\perp)$, $\beta(\vec{k},\Omega,\textbf{K}_\perp)$ are those in \cite{RevModPhys.80.787}.

Also we want to point out that the massive 3+1 case is very similar to the massless 3+1 case, since both Eqs. (\ref{vacuum_state_property_1_3}, \ref{ordering_Rindler_1_3}) hold and the only difference relies on the Bogolyubov coefficients, which read \cite{RevModPhys.80.787}
\begin{subequations} \label{Bogolyubov_coefficients_1_3}
\begin{equation}
\alpha(\vec{k},\Omega,\textbf{K}_\perp) = \frac{\delta^2(\textbf{k}_\perp-\textbf{K}_\perp)\exp \left(\frac{\beta \Omega}{4} \right)}{\sqrt{4 \pi a \omega(k) \sinh \left( \frac{\beta \Omega}{2} \right)}} \left( \frac{\omega(k)+k_x}{\omega(k)-k_x} \right)^{\frac{i\Omega}{2a}},
\end{equation}
\begin{equation}
\beta(\vec{k},\Omega,\textbf{K}_\perp) = \exp \left( \frac{\beta \Omega}{2} \right) \alpha(\vec{k},\Omega,-\textbf{K}_\perp),
\end{equation}
\end{subequations}
where $\vec{k}=(k_x,\textbf{k}_\perp)=(k_x,k_y,k_z)$, $\omega^2(k) = m^2 + k^2$ and $m$ is the mass of the field. The massless case can be restored by simply imposing $\omega(k) = |k|$ in Eqs. (\ref{Bogolyubov_coefficients_1_3}).

By following again the steps of Sec. \ref{Method} while carefully taking track of the change in sign for the transverse momenta $\textbf{K}_\perp$, one can prove that the results of Sec. \ref{Result} are still valid, with the only difference given by the new Bogolyubov coefficients [Eqs. (\ref{Bogolyubov_coefficients_1_3})]:
\begin{subequations}
\begin{align} \label{chi_bar_final_1_3}
 & \bar{\chi}^{(p)}(\mathbf{U}, \mathbf{U}')[\xi, \xi^*] = \chi^{(p)}_0[\xi,\xi^*] \sum_{\substack{\mathbf{S} \subseteq \mathbf{U} \\ \mathbf{S}' \subseteq \mathbf{U}' }} c(\mathbf{U} \setminus \mathbf{S},\mathbf{U}' \setminus \mathbf{S}')  \nonumber \\
 & \times \prod_{\vec{k}\in\mathbf{S}}  \{ - L (\vec{k}) [\xi,\xi^*]\}^*  \prod_{\vec{k}'\in\mathbf{S}'}L (\vec{k}') [\xi,\xi^*]  ,
\end{align}
\begin{equation} \label{c_combinatory_1_3}
c \left( \{\vec{k}_i\}_{i=1}^M,\{\vec{k}'_{j}\}_{j=1}^N \right) = \delta_{MN} \sum_\mathcal{P} \prod_{i} \delta^3(\vec{k}_i - \vec{k}'_{\mathcal{P}(i)})
\end{equation}
\begin{align}
L (\vec{k}) [\xi,\xi^*] = & \int_{0}^{\infty} d\Omega \int_{\mathbb{R}^2} d^2\textbf{K}_\perp \left[\alpha^*(\vec{k},\Omega,\textbf{K}_\perp) \xi(\Omega,\textbf{K}_\perp) \right. \nonumber \\
& \left. - \beta(\vec{k},\Omega,\textbf{K}_\perp) \xi^*(\Omega,\textbf{K}_\perp) \right],
\end{align}
\begin{align}
\chi^{(p)}_0[\xi,\xi^*] = & \exp \left( \int_{0}^{\infty} d\Omega \int_{\mathbb{R}^2} d^2\textbf{K}_\perp |\xi(\Omega,\textbf{K}_\perp)|^2 \right. \nonumber \\
& \left. \times \left[- n_0(\Omega) + \frac{p-1}{2} \right] \right).
\end{align}
\end{subequations}

\section{Single-particle state}\label{Single-particle state}

In the present section, we focus on single-particle states as the simplest example of Minkowski-Fock states. For such states, we derive the characteristic function in the accelerated frame using Eqs. (\ref{chi_final}). As a practical application of $\chi^{(p)}[\xi, \xi^*]$, we derive mean values of $\hat{\rho}$ related to the probability distribution to find a Rindler particle with a specific momentum and the correlation between Rindler particles with different momenta. We show how the probability distribution changes from the inertial to the accelerated frame. Moreover we show how the correlation between different momenta has the same expression for both the inertial to the accelerated frame if we exclude the vacuum background. Finally, we consider Gaussian wave-functions and plot the results in the limit of well-localized wave-packets in the momentum space.

We start from the definition of Minkowski-Fock single-particle states
\begin{equation} \label{single_particle}
| \psi \rangle = \int_{-\infty}^{+\infty} dk \tilde{\psi}(k) \hat{a}^\dagger (k) | 0_M \rangle,
\end{equation}
with $\tilde{\psi}(k)$ as a normalized wave-function
\begin{equation} \label{single_particle_normalization}
\int_{-\infty}^{+\infty} dk \left| \tilde{\psi}(k) \right|^2 = 1.
\end{equation}
For such state, Eq. (\ref{chi_final_a}) reads
\begin{align} \label{single_particle_chi}
\chi^{(p)}[\xi, \xi^*] = & \int_{-\infty}^{+\infty} dk \tilde{\psi}(k) \int_{-\infty}^{+\infty} dk' \tilde{\psi}^*(k') \nonumber \\
& \times \bar{\chi}^{(p)} (\{k\},\{k'\}) [\xi, \xi^*].
\end{align}
Therefore, in this case, it is sufficient to show the explicit form of $ \bar{\chi}^{(p)}(\{k\},\{k'\})[\xi, \xi^*]$, which has the following diagrammatic expression
\begin{equation} \label{single_particle_diagram}
\bar{\chi}^{(p)}(\{k\},\{k'\})[\xi, \xi^*] = k \text{ \textcolor{white}{---} } k' + k \text{ --- } k'.
\end{equation}
The explicit expression for Eq. (\ref{single_particle_diagram}) is
\begin{align} \label{single_particle_chi_bar}
\bar{\chi}^{(p)}(\{k\},\{k'\})[\xi, \xi^*] = & \chi^{(p)}_0[\xi,\xi^*] \delta(k-k') - \chi^{(p)}_0[\xi,\xi^*] \nonumber \\
& \times \{L (k) [\xi,\xi^*] \}^* L (k') [\xi,\xi^*].
\end{align}

Eqs. (\ref{single_particle_normalization}, \ref{single_particle_chi}, \ref{single_particle_chi_bar}) lead to
\begin{align} \label{single_particle_chi_final}
\chi^{(p)}[\xi, \xi^*] = & \chi^{(p)}_0[\xi,\xi^*] \nonumber \\
& -  \chi^{(p)}_0[\xi,\xi^*] \left| \int_{-\infty}^{+\infty} dk \tilde{\psi}(k) \{ L (k) [\xi,\xi^*] \}^* \right|^2.
\end{align}

Eq. (\ref{single_particle_chi_final}) can be used in order to extract the probability density
\begin{equation}\label{probability_density}
\left\langle \hat{n}_R(K) \right\rangle_{\hat{\rho}} = \text{Tr} \left( \hat{\rho} \hat{b}^\dagger_R(K) \hat{b}_R(K) \right).
\end{equation}
Indeed, the left side of Eq. (\ref{probability_density}) can be obtained through the following derivatives:
\begin{equation}
\left\langle \hat{n}_R(K) \right\rangle_{\hat{\rho}} = - \left. \frac{\delta}{\delta \xi(K)} \frac{\delta}{\delta \xi^*(K)} \chi^{(1)}[\xi, \xi^*] \right|_{\xi = 0}.
\end{equation}
The result is the following
\begin{align}
\left\langle \hat{n}_R(K) \right\rangle_{\hat{\rho}} = & \left\langle \hat{n}_R(K) \right\rangle_{\hat{\rho}_0} + n_R(K),
\end{align}
where
\begin{align}
n_R(K) = & \left| \int_{-\infty}^{+\infty} dk \tilde{\psi}(k) \alpha(k,K) \right|^2 \nonumber \\
& + \left| \int_{-\infty}^{+\infty} dk \tilde{\psi}(k) \beta^*(k,K) \right|^2
\end{align}
represents the probability distribution to find a particle with momentum $K$ over the vacuum background.

$\left\langle \hat{n}_R(K) \right\rangle_{\hat{\rho}}$ of Eq. (\ref{probability_density}) can be compared with the Minkowski probability density
\begin{equation}\label{Minkowski_probability_density}
\left\langle \hat{n}_M(k) \right\rangle_{|\psi\rangle \langle \psi |} = \text{Tr} \left( |\psi\rangle \langle \psi | \hat{a}^\dagger(k) \hat{a}(k) \right).
\end{equation}
In the Minkowski space-time, the probability distribution to find a particle with momentum $k$ over the vacuum background $n_M(k)$ is obviously directly identified with $\left\langle \hat{n}_M(k) \right\rangle_{|\psi\rangle \langle \psi |}$. Moreover, for single-particle states, it is well known that
\begin{equation}
n_M(k) = \left| \tilde{\psi}(k) \right|^2.
\end{equation}

Differently from $\left\langle \hat{n}_R(K) \right\rangle_{\hat{\rho}}$, a quantity that can be obtained through more than two derivatives of the characteristic function is the following:
\begin{align}
& \left\langle \hat{n}_R(K) \hat{n}_R(K') \right\rangle_{\hat{\rho}} = \delta(K-K') \left\langle \hat{n}_R(K) \right\rangle_{\hat{\rho}} \nonumber \\
& + \left. \frac{\delta}{\delta \xi(K)} \frac{\delta}{\delta \xi(K')} \frac{\delta}{\delta \xi^*(K)}  \frac{\delta}{\delta \xi^*(K')} \chi^{(1)}[\xi, \xi^*] \right|_{\xi = 0}.
\end{align}
From $\left\langle \hat{n}_R(K) \right\rangle_{\hat{\rho}}$ and $\left\langle \hat{n}_R(K) \hat{n}_R(K') \right\rangle_{\hat{\rho}}$ it is possible to derive a quantity that can measure correlations between particles with momentum $K$ and $K'$ over the vacuum background:
\begin{align} \label{C_R}
& C_R (K,K') = \left\langle \hat{n}_R(K) \hat{n}_R(K') \right\rangle_{\hat{\rho}} -  \left\langle \hat{n}_R(K)  \right\rangle_{\hat{\rho}} \left\langle \hat{n}_R(K') \right\rangle_{\hat{\rho}} \nonumber \\
&  - [\left\langle \hat{n}_R(K) \hat{n}_R(K') \right\rangle_{\hat{\rho}_0}  - \left\langle \hat{n}_R(K) \right\rangle_{\hat{\rho}_0} \left\langle \hat{n}_R(K') \right\rangle_{\hat{\rho}_0}].
\end{align}
In the Minkowski case such quantity reads
\begin{align} \label{C_M}
C_M (k,k') = & \left\langle \hat{n}_M(k) \hat{n}_M(k') \right\rangle_{|\psi\rangle \langle \psi |} \nonumber \\
&   - \left\langle \hat{n}_M(k)  \right\rangle_{|\psi\rangle \langle \psi |}\left\langle \hat{n}_M(k') \right\rangle_{|\psi\rangle \langle \psi |}.
\end{align}

In the case of single-particle states Eqs. (\ref{C_R}, \ref{C_M}) lead to
\begin{subequations}\label{single_particle_C}
\begin{align}\label{single_particle_C_R}
C_R (K,K') = & \delta(K-K')n_R(K) [ 1 + 2 n_0(K)] \nonumber \\
&  - n_R(K) n_R(K').
\end{align}
\begin{equation}
C_M (k,k') = \delta(k-k') n_M(k) - n_M(k) n_M(k').
\end{equation}
\end{subequations}
From Eq. (\ref{single_particle_C}) we can conclude that for different momenta -- i.e. $K \neq K'$ and $k \neq k'$ -- the form of $C_R (K,K')$ and $C_M (k,k')$ are the same. This means that, besides the vacuum background, no correlation has been introduced by shifting from the Minkowki to the Rindler frame.

\begin{figure}[h]
\includegraphics[]{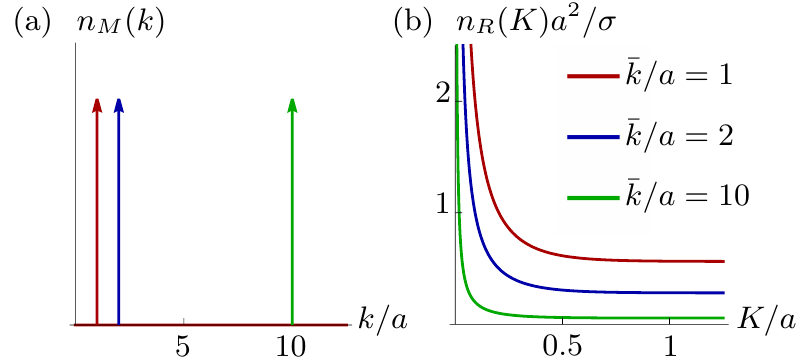}
\caption{Representation of how the probability density of a particle with fixed momentum $\bar{k}$ changes with the acceleration. In panel a, the distribution of $n_M(k)$ is shown for different values of $\bar{k}$. For the same values of $\bar{k}$, function $n_R(K)$ defined by the right side of Eq. (\ref{n_R}) is shown in panel b.}\label{f_figure}
\end{figure}

In order to give a practical application, let us consider a Gaussian wave-function $\tilde{\psi}(k) = G(k;\bar{k},\sigma)$, with
\begin{equation} \label{Gaussian_psi_k}
G(k;\bar{k},\sigma) = \frac{1}{\sqrt[4]{\pi}\sqrt{\sigma}} \exp \left( - \frac{(k-\bar{k})^2}{2 \sigma^2} \right)
\end{equation}
and let us consider a well-localized wave-packet -- i.e. $\sigma/a \rightarrow 0$ -- then the leading terms of $n_R(K)$ is
\begin{equation} \label{n_R}
n_R(K) \approx \frac{\sigma}{a} \frac{\theta(\bar{k} K)}{\sqrt{\pi} |\bar{k}|} \coth\left( \frac{\beta}{2}|K| \right).
\end{equation}
On the other hand the Minkowski probability density has the usual distributional limit
\begin{equation}
n_M(k) \approx \delta(k-\bar{k}).
\end{equation}
In Fig. \ref{f_figure}, we plot the transformation from $n_M(k)$ to $n_R(K)$ for some single-particle states with different $\bar{k}$.

\section{Single-particle state in 3+1 dimensions} \label{Single-particle_state_in_3+1_dimensions}

At this point it is worth to mention the fact that only in the 1+1-dimensional massless case, a direct comparison between $C_R(K,K')$ and $C_M (k,k')$ in terms of particle momenta is possible. Indeed, in the 3+1-dimensional case, Rindler particles are identified through the energies $\Omega$, $\Omega'$ and transverse momenta $\textbf{K}_\perp$, $\textbf{K}'_\perp$, while Minkowksi particles still with momenta components $\vec{k}$, $\vec{k}'$. In that case, a more reasonable comparison must be made between $C_R(\Omega, \textbf{K}_\perp, \Omega', \textbf{K}'_\perp)$ and $C_M$ as a function of the energies $\omega(k) = |\vec{k}|$, $\omega(k')$ and transverse momenta $\textbf{k}_\perp$, $\textbf{k}'_\perp$. However such function is not well defined since the sign of $k_x$ and $k'_x$ represents a further degeneracy for the energy states. Let us, therefore, choose $k_x>0$, $k'_x>0$ and define in this way $C_M(\omega, \textbf{k}_\perp, \omega', \textbf{k}'_\perp)$.

By following the results of Sec. \ref{1_3_massive_field}, it can be shown that the form of Eq. (\ref{single_particle_C}) remains the same for $C_R(\Omega, \textbf{K}_\perp, \Omega', \textbf{K}'_\perp)$ and $C_M(\omega, \textbf{k}_\perp, \omega', \textbf{k}'_\perp)$, with the only difference given by the fact that $\Omega$ and $\omega$ are positive and the explicit value of $n_R(\Omega,\textbf{K}_\perp)$ changes because of the new Bogolyubov coefficients:
\begin{subequations}
\begin{align}
& C_R(\Omega, \textbf{K}_\perp, \Omega', \textbf{K}'_\perp) = \delta(\Omega-\Omega') \delta^2(\textbf{K}_\perp - \textbf{K}'_\perp)  \nonumber \\
& \times n_R(\Omega,\textbf{K}_\perp) [ 1 + 2 n_0(\Omega)] - n_R(\Omega,\textbf{K}_\perp) n_R(\Omega',\textbf{K}'_\perp),
\end{align}
\begin{align}
C_M(\omega, \textbf{k}_\perp, \omega', \textbf{k}'_\perp) =&  \delta(\omega-\omega') \delta^2(\textbf{k}_\perp - \textbf{k}'_\perp) n_M(\omega, \textbf{k}_\perp) \nonumber \\
& - n_M(\omega, \textbf{k}_\perp) n_M(\omega', \textbf{k}'_\perp),
\end{align}
\begin{align}
n_R(\Omega,\textbf{K}_\perp) = & \left| \int_{\mathbb{R}^3} d^3k \tilde{\psi}(\vec{k}) \alpha(\vec{k},\Omega,\textbf{K}_\perp) \right|^2 \nonumber \\
& + \left| \int_{\mathbb{R}^3} d^3k \tilde{\psi}(\vec{k}) \beta^*(\vec{k},\Omega,\textbf{K}_\perp) \right|^2.
\end{align}
\end{subequations}
We can state that even for the 3+1-dimesional case, no correlation is introduced in the Rindler frame. However, one must be careful when considering the different spectra in the two frames, since in the Minkowski case, a two-degeneracy for states with fixed energy and transverse momentum exists.

Lastly, we want to mention the 3+1-dimesional massive case. As shown by Sec. \ref{1_3_massive_field}, no difference occurs from the 3+1-dimesional massless case, with the exception given by the mass-dependent Bogolyubov coefficients that result in a different explicit value for $n_R(\Omega,\textbf{K}_\perp)$. However, one has to carefully consider the different spectra in the two frames. Indeed, while $\Omega$ is still defined from $0$ to $\infty$, $\omega$ is defined for $\omega>m$. Therefore, such difference in the particle spectra must be taken into account for a real comparison between $C_M(\omega, \textbf{k}_\perp, \omega', \textbf{k}'_\perp)$ and $C_R(\Omega, \textbf{K}_\perp, \Omega', \textbf{K}'_\perp)$.

\section{Two-particle state}\label{Two-particle state}

In this section we consider two-particle states. Differently from the one-particle states, we show that $C_R (K,K')$ and $C_M (k,k')$ have different forms. Moreover, we provide the example of Gaussian-like wave-functions and consider again the limit of well-localized wave-packets in the momentum space.

A Minkowski two-particle state can be defined through a wave-function $\psi(k,k')$ which is symmetric with respect to a switch between $k$ and $k'$:
\begin{equation}
| \psi \rangle = \int_{-\infty}^{+\infty} dk \int_{-\infty}^{+\infty} dk'  \tilde{\psi}(k,k') \hat{a}^\dagger (k) \hat{a}^\dagger (k') | 0_M \rangle,
\end{equation}
with
\begin{subequations}
\begin{equation} \label{two_particle_symmetry}
 \tilde{\psi}(k,k') =  \tilde{\psi}(k',k),
\end{equation}
\begin{equation} \label{two_particle_normalization}
\int_{-\infty}^{+\infty} dk \int_{-\infty}^{+\infty} dk' \left| \tilde{\psi}(k,k') \right|^2 = \frac{1}{2}.
\end{equation}
\end{subequations}
For such state, Eq. (\ref{chi_final_a}) reads
\begin{widetext}
\begin{equation} \label{two_particle_chi}
\chi^{(p)}[\xi, \xi^*] = \int_{-\infty}^{+\infty} dk_1 \int_{-\infty}^{+\infty} dk_2 \tilde{\psi}(k_1,k_2) \int_{-\infty}^{+\infty} dk'_1  \int_{-\infty}^{+\infty} dk'_2 \tilde{\psi}^*(k'_1,k'_2) \bar{\chi}^{(p)} (\{k_1,k_2\},\{k'_1,k'_2\}) [\xi, \xi^*].
\end{equation}
$ \bar{\chi}^{(p)} (\{k_1,k_2\},\{k'_1,k'_2\}) [\xi, \xi^*]$ has the following diagrammatic expression
\begin{equation} \label{two_particles_diagram}
\begin{tikzpicture}
\node (0_1) [] {$ \bar{\chi}^{(p)}(\{k_1,k_2\},\{k'_1,k'_2\})[\xi, \xi^*] = $};

\node (1L_vdots) [right=0cm of 0_1] {};
\node (1R_vdots) [right of=1L_vdots] {};

\node (1L_k_1) [above=0cm of 1L_vdots] {$k_1$};
\node (1L_k_2) [below=0cm of 1L_vdots] {$k_2$};
\node (1R_k_1) [above=0cm of 1R_vdots] {$k'_1$};
\node (1R_k_2) [below=0cm of 1R_vdots] {$k'_2$};

\node (1_2) [right=0cm of 1R_vdots] {$+$};
\node (2L_vdots) [right=0cm of 1_2] {};
\node (2R_vdots) [right of=2L_vdots] {};

\node (2L_k_1) [above=0cm of 2L_vdots] {$k_1$};
\node (2L_k_2) [below=0cm of 2L_vdots] {$k_2$};
\node (2R_k_1) [above=0cm of 2R_vdots] {$k'_1$};
\node (2R_k_2) [below=0cm of 2R_vdots] {$k'_2$};

\node (2_3) [right=0cm of 2R_vdots] {$+$};
\node (3L_vdots) [right=0cm of 2_3] {};
\node (3R_vdots) [right of=3L_vdots] {};

\node (3L_k_1) [above=0cm of 3L_vdots] {$k_1$};
\node (3L_k_2) [below=0cm of 3L_vdots] {$k_2$};
\node (3R_k_1) [above=0cm of 3R_vdots] {$k'_1$};
\node (3R_k_2) [below=0cm of 3R_vdots] {$k'_2$};

\node (3_4) [right=0cm of 3R_vdots] {$+$};
\node (4L_vdots) [right=0cm of 3_4] {};
\node (4R_vdots) [right of=4L_vdots] {};

\node (4L_k_1) [above=0cm of 4L_vdots] {$k_1$};
\node (4L_k_2) [below=0cm of 4L_vdots] {$k_2$};
\node (4R_k_1) [above=0cm of 4R_vdots] {$k'_1$};
\node (4R_k_2) [below=0cm of 4R_vdots] {$k'_2$};

\node (4_5) [right=0cm of 4R_vdots] {$+$};
\node (5L_vdots) [right=0cm of 4_5] {};
\node (5R_vdots) [right of=5L_vdots] {};

\node (5L_k_1) [above=0cm of 5L_vdots] {$k_1$};
\node (5L_k_2) [below=0cm of 5L_vdots] {$k_2$};
\node (5R_k_1) [above=0cm of 5R_vdots] {$k'_1$};
\node (5R_k_2) [below=0cm of 5R_vdots] {$k'_2$};

\node (5_6) [right=0cm of 5R_vdots] {$+$};
\node (6L_vdots) [right=0cm of 5_6] {};
\node (6R_vdots) [right of=6L_vdots] {};

\node (6L_k_1) [above=0cm of 6L_vdots] {$k_1$};
\node (6L_k_2) [below=0cm of 6L_vdots] {$k_2$};
\node (6R_k_1) [above=0cm of 6R_vdots] {$k'_1$};
\node (6R_k_2) [below=0cm of 6R_vdots] {$k'_2$};

\node (6_7) [right=0cm of 6R_vdots] {$+$};
\node (7L_vdots) [right=0cm of 6_7] {};
\node (7R_vdots) [right of=7L_vdots] {};

\node (7L_k_1) [above=0cm of 7L_vdots] {$k_1$};
\node (7L_k_2) [below=0cm of 7L_vdots] {$k_2$};
\node (7R_k_1) [above=0cm of 7R_vdots] {$k'_1$};
\node (7R_k_2) [below=0cm of 7R_vdots] {$k'_2$};

\draw[-] (2L_k_1) to (2R_k_1);
\draw[-] (3L_k_2) to (3R_k_2);
\draw[-] (4L_k_1) to (4R_k_2);
\draw[-] (5L_k_2) to (5R_k_1);
\draw[-] (6L_k_1) to (6R_k_1);
\draw[-] (6L_k_2) to (6R_k_2);
\draw[-] (7L_k_1) to (7R_k_2);
\draw[-] (7L_k_2) to (7R_k_1);

\end{tikzpicture}
\end{equation}

By using the symmetry (\ref{two_particle_symmetry}) and the normalization (\ref{two_particle_normalization}), we can write Eq. (\ref{two_particle_chi}) in the following way
\begin{align} \label{two_particle_chi_final}
\chi^{(p)}[\xi, \xi^*] = &  \left\lbrace \left| \int_{-\infty}^{+\infty} dk \int_{-\infty}^{+\infty} dk' \tilde{\psi}(k,k')  L^*(k) [\xi,\xi^*]  L^*(k') [\xi,\xi^*] \right|^2 - 4 \int_{-\infty}^{+\infty} dk \left| \int_{-\infty}^{+\infty} dk' \tilde{\psi}(k,k')  L^*(k') [\xi,\xi^*] \right|^2 + 1 \right\rbrace \nonumber \\
& \times \chi^{(p)}_0[\xi, \xi^*].
\end{align}

Eq. (\ref{two_particle_chi_final}) can be used in order to extract $n_R(K)$ and $C_R(K,K')$:
\begin{subequations}
\begin{equation}
n_R(K) =  4 \int_{-\infty}^{+\infty} dk  \left[ \left| \int_{-\infty}^{+\infty} dk' \tilde{\psi}(k,k')  \alpha(k',K) \right|^2 + \left| \int_{-\infty}^{+\infty} dk' \tilde{\psi}(k,k')  \beta^*(k',K)\right|^2 \right]
\end{equation}
\begin{align}\label{two_particle_C_R}
C_R(K,K') = &  \delta(K-K')n_R(K) [ 1 + 2 n_0(K)]  - n_R(K) n_R(K')  \nonumber \\
& + 4 \left| \int_{-\infty}^{+\infty} dk \int_{-\infty}^{+\infty} dk' \tilde{\psi}(k,k')  \alpha(k,K)\alpha(k',K') \right|^2  + 4 \left| \int_{-\infty}^{+\infty} dk \int_{-\infty}^{+\infty} dk' \tilde{\psi}(k,k')  \beta^*(k,K)\beta^*(k',K') \right|^2 \nonumber\\
& + 4 \int_{-\infty}^{+\infty} dk_1 \int_{-\infty}^{+\infty} dk_1' \tilde{\psi}(k_1,k_1') \int_{-\infty}^{+\infty} dk_2  \int_{-\infty}^{+\infty} dk'_2 \tilde{\psi}^*(k_2,k'_2) \nonumber\\
& \times [ \alpha(k_1,K')\beta(k_2,K) + \beta(k_2,K')\alpha(k_1,K) ]  [\beta^*(k_1',K')\alpha^*(k_2',K) + \alpha^*(k_2',K')\beta^*(k_1',K)]
\end{align}
\end{subequations}
\end{widetext}

\begin{figure}[h]
\includegraphics[]{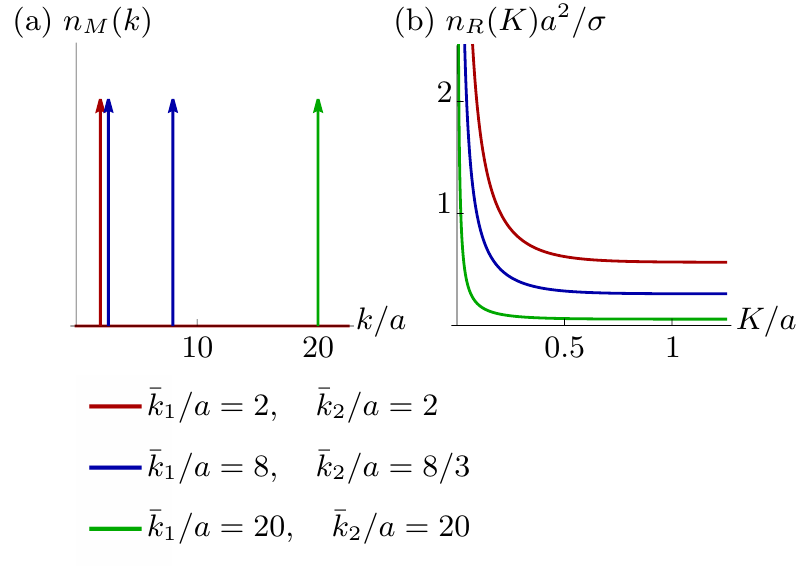}
\caption{Representation of how the probability density of two particles with fixed momenta $\bar{k}_1$ and $\bar{k}_2$ changes with the acceleration. In panel a, the distribution of $n_M(k)$ is shown for different values of $\bar{k}_1$ and $\bar{k}_2$. For the same values of $\bar{k}_1$ and $\bar{k}_2$, function $n_R(K)$ defined by the right side of Eq. (\ref{n_R_two}) is shown in panel b.}\label{f_2_figure}
\end{figure}

\begin{figure}[h]
\includegraphics[]{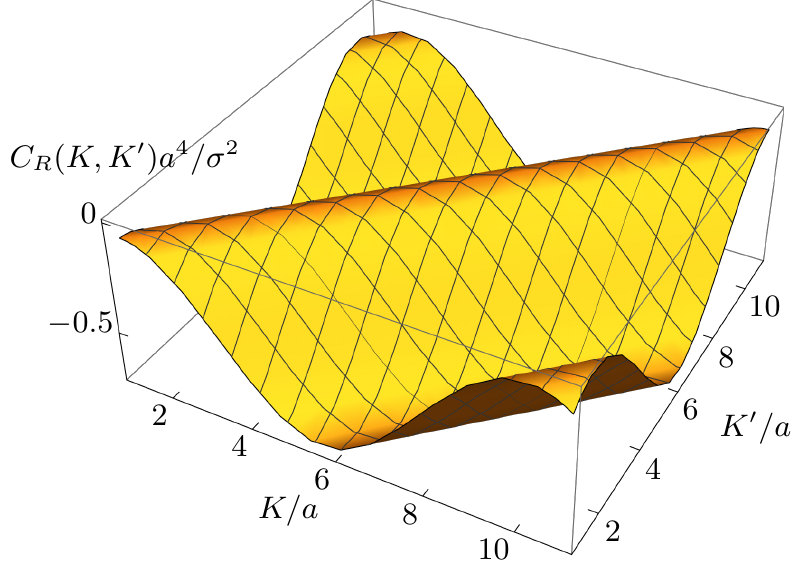}
\caption{In the present figure we show the correlation $C_R(K,K')$ in the Rindler space-time between particles with different momenta $K \neq K'$ as defined by the right side of Eq. (\ref{C_R_two}). The solution we have chosen is with $\bar{k}_1/a = 1$ and $\bar{k}_2/a = 2$.}\label{C_figure}
\end{figure}

On the other hand, in the Minkowski case,
\begin{subequations}
\begin{equation}
n_M(k) =  4 \int_{-\infty}^{+\infty} dk'   \left| \tilde{\psi}(k,k')  \right|^2 
\end{equation}
\begin{align}
C_M(k,k') = &  \delta(k-k')  n_M(k)  - n_M(k) n_M(k') \nonumber \\
& +  4 \left|  \tilde{\psi}(k,k') \right|^2
\end{align}
\end{subequations}
It is interesting to notice how in this case the form of $C_R (K,K')$ and $C_M (K,K')$ are different. This means that a correlation between particles with different momenta has been introduced by shifting from the inertial to the accelerated frame. The result we have obtained for the two-particle states differs from the single-particle state, as we have seen in Eqs. (\ref{single_particle_C}). The consequence is that we can actually discriminate Minkowski single-particle states from two-particle states in the Rindler space-time by looking at non-diagonal values of $C_R (K,K')$.

Finally, we want to provide Gaussian wave-function as practical applications for the theory and focus on the limit of well-localized wave-packets in the momentum space. If we define
\begin{equation}
\tilde{\psi}(k,k') \propto G(k;\bar{k}_1,\sigma)G(k';\bar{k}_2,\sigma) + G(k;\bar{k}_2,\sigma)G(k';\bar{k}_1,\sigma),
\end{equation}
we obtain the following results
\begin{widetext}
\begin{subequations}
\begin{equation} \label{n_R_two}
n_R(K) \approx \frac{\sigma}{a} \frac{1}{\sqrt{\pi}} \left[ \frac{\theta(\bar{k}_1 K)}{|\bar{k}_1|} + \frac{\theta(\bar{k}_2 K)}{|\bar{k}_2|} \right] \coth\left( \frac{\beta}{2}|K| \right),
\end{equation}
\begin{equation}
n_M(k) \approx \delta (k-\bar{k}_1) + \delta (k-\bar{k}_2)
\end{equation}
\begin{align} \label{C_R_two}
C_R(K,K') \approx &   \delta(K-K') \frac{\sigma}{a} \frac{1}{\sqrt{\pi}}  [ 1 + 2 n_0(K)]  \left[ \frac{\theta(\bar{k}_1 K)}{|\bar{k}_1|} + \frac{\theta(\bar{k}_2 K)}{|\bar{k}_2|} \right] \coth\left( \frac{\beta}{2}|K| \right) \nonumber \\
& + \left( \frac{\sigma}{a} \right)^2 \frac{1}{\pi} \text{csch} \left( \frac{\beta}{2}|K| \right) \text{csch} \left( \frac{\beta}{2}|K'| \right)  \left\lbrace   \frac{\theta(\bar{k}_1 K) \theta(\bar{k}_2 K') + \theta(\bar{k}_2 K) \theta(\bar{k}_1 K')}{ 2 | \bar{k}_1 \bar{k}_2 |}  \right. \nonumber \\
& - \left[ \frac{\theta(\bar{k}_1 K)\theta(\bar{k}_1 K')}{|\bar{k}_1|^2} + \frac{\theta(\bar{k}_2 K)\theta(\bar{k}_2 K')}{|\bar{k}_2|^2} \right]  \cosh\left( \frac{\beta}{2}|K| \right) \cosh\left( \frac{\beta}{2}|K'| \right)  \nonumber \\
& +  \left. \frac{\theta(\bar{k}_1 K) \theta(\bar{k}_1 K')\theta(\bar{k}_1 \bar{k}_2)}{ | \bar{k}_1 \bar{k}_2 |} \cos \left( \frac{K-K'}{a} \ln \left| \frac{\bar{k}_1}{\bar{k}_2} \right|  \right)  \left[ 2 \cosh \left( \frac{\beta}{2}|K| \right) \cosh \left( \frac{\beta}{2}|K'| \right)  + 1 \right]  \right\rbrace
\end{align}
\begin{equation}\label{C_M_two}
C_M(k,k') = o ( (\sigma/a)^2 )
\end{equation}
\end{subequations}
\end{widetext}
when $\sigma/a \rightarrow 0$.

It is possible to notice that for any choice of $\bar{k}_1$ and $\bar{k}_2$ with the same sign, Eq. (\ref{n_R_two}) has the same form of Eq. (\ref{n_R}) for a specific choice of $\bar{k}$:
\begin{equation}
\frac{1}{\bar{k}} = \frac{1}{\bar{k}_1} + \frac{1}{\bar{k}_2}.
\end{equation}
This result can be observed by comparing Fig. \ref{f_figure} with Fig. \ref{f_2_figure}, where for specific choices of $\bar{k}$, $\bar{k}_1$ and $\bar{k}_2$, we have been able to reproduce the same probability density in the Rindler space-time even when the number of particles differ. This means that if we look at right-Rindler particle density distribution, Minkowski two-particle states become indistinguishable from single-particles with momentum equal to half of the harmonic mean of the two-particle momenta. For instance, if both particles have the same momentum $\bar{k}_1 = \bar{k}_2$, the two-particle state becomes indistinguishable from a single-particle with momentum $\bar{k} = 2 \bar{k}_1$. The same result does not hold for the inertial observer, who is actually able to distinguish the two cases -- e.g.: by integrating $n_M(k)$ with respect to $k$ and obtaining $1$ for single-particles and $2$ for two-particles. This result point towards the possibility for the observer to discriminate between the inertial to the accelerated frame.

While single-particle and two-particle states cannot be distinguished by the expression of $n_R(K)$, the correlation function $C_R(K,K')$ offers a way to discriminate between the two of them. Indeed, Eqs. (\ref{single_particle_C_R}, \ref{C_R_two}) do not have the same form. While in the case of single-particle states the form of $C_R(K,K')$ is the same of $C_M(K,K')$, in the two particle state they differ. Specifically, while $C_M(K,K')$ vanishes faster than $(\sigma/a)^2$, $C_R(K,K')$ has a specific distribution of order $(\sigma/a)^2$ shown in Fig. \ref{C_figure}. This also provides a way to discriminate between the Minkowski to the right-Rindler frame.

\section{Conclusions}\label{Conclusions}

The method we have adopted allowed us to extract the general expression of $\chi^{(p)}[\xi, \xi^*]$ and any derivable mean values for Minkowki-Fock states. To the best of our knowledge, this result is not known in literature. Moreover, it allowed us to investigate how quantities such as $n_M(k)$ and $C_M(k,k')$ transform from an inertial to an accelerated observer. Specifically, we have extracted $n_R(k)$ and $C_R(k,k')$ for single-particle and two-particle states. An interesting result of such analysis is that $n_R(K)$, differently from $n_M(k)$, cannot be used as a general way to detect the presence of a second Minkowski particle. On the other hand, by measuring both $C_R(k,k')$ and $n_R(k)$, one can distinguish between single-particle and two-particle states in the accelerated frame, since in the former case $C_R(k,k')$ has the same form of $C_M(k,k')$, while in the letter they differ. A remarkable outcome is that a non-inertial observer that is able to generate two independent particles with different momenta will measure fictitious correlations dependent on the acceleration. This opens a way to test non-inertial quantum field theory.

\appendix

\section{A proof for Eq. (\ref{chi_final})}\label{A_proof_for}

In the present section, we want to show a proof for Eq. (\ref{chi_final}), through the procedure described in the main paper and through the use of the following identities for the Bogolyubov coefficients
\begin{equation} \label{alpha_to_beta}
\alpha(k,K) = \exp \left( \frac{\beta}{2} |K| \right) \beta(k,K) ,
\end{equation}
\begin{align} \label{identity_K}
& \int_{-\infty}^{+\infty}dK 2 \sinh \left( \frac{\beta}{2} |K| \right) [ \alpha(k,K)\beta^*(k',K) \nonumber \\
& + \beta^*(k,K)\alpha(k',K)  ] = \delta(k-k').
\end{align}
Eq. (\ref{alpha_to_beta}) can be extracted from the following identity
\begin{equation} \label{alpha_beta_to_alpha_beta_proof}
F(k,K) = \exp \left( \text{sign}(k) \frac{\beta}{2} K \right) F(-k,K),
\end{equation}
while Eq. (\ref{identity_K}) can be proven by the following chain of identities:
\begin{widetext}
\begin{align}
& \int_{-\infty}^{+\infty}dK \theta(kK) \theta(k'K) 2 \sinh \left( \frac{\beta}{2} |K| \right) \frac{|K|}{\sqrt{kk'}}  \left[ F(k,K)F(k',-K) + F(k,-K)F(k',K)  \right] \nonumber \\
= &  \frac{\theta(kk')}{\sqrt{kk'}} \left[ \int_{-\infty}^{+\infty}dK \theta(kK) 2 \sinh \left( \frac{\beta}{2} |K| \right) |K|  F(k,K)F(k',-K) \right. \nonumber \\
& \left. + \int_{-\infty}^{+\infty}dK \theta(kK) 2 \sinh \left( \frac{\beta}{2} |K| \right) |K| F(k,-K)F(k',K)  \right] \nonumber \\
= &  \frac{\theta(kk')}{\sqrt{kk'}} \left[ \int_{-\infty}^{+\infty}dK \theta(kK) 2 \sinh \left( \frac{\beta}{2} |K| \right) |K|  F(k,K)F(k',-K) \right. \nonumber \\
& \left. + \int_{-\infty}^{+\infty}dK \theta(-kK) 2 \sinh \left( \frac{\beta}{2} |K| \right) |K| F(k,K)F(k',-K)  \right] \nonumber \\
= &  \frac{\theta(kk')}{\sqrt{kk'}} \int_{-\infty}^{+\infty}dK \left[ \theta(kK)+\theta(-kK) \right] 2 \sinh \left( \frac{\beta}{2} |K| \right) |K|  F(k,K)F(k',-K) \nonumber \\
= &  \frac{\theta(kk')}{\sqrt{kk'}} \int_{-\infty}^{+\infty}dK 2 \sinh \left( \frac{\beta}{2} |K| \right) |K|  F(k,K)F(k',-K) \nonumber \\
= & \frac{\theta(kk')}{\sqrt{kk'}} \int_{-\infty}^{+\infty}dK 2 \sinh \left( \frac{\beta}{2} |K| \right) \frac{|K|}{(2 \pi a)^2} \left| \Gamma \left( \frac{i K}{a} \right) \right|^2 \exp \left( i \frac{K}{a} \ln \left| \frac{k}{k'} \right| \right) \nonumber \\
= & \frac{\theta(kk')}{\sqrt{kk'}} \int_{-\infty}^{+\infty} \frac{dK}{2 \pi a}  \exp \left( i \frac{K}{a} \ln \left| \frac{k}{k'} \right| \right)\nonumber \\
= & \frac{\theta(kk')}{\sqrt{kk'}} \delta \left( \ln \left| \frac{k}{k'} \right| \right) \nonumber \\
= & \frac{\theta(kk')}{|k|} \delta \left( \ln \left| \frac{k}{k'} \right| \right) \nonumber \\
= & \theta(kk') \delta \left( |k| - |k'| \right)\nonumber \\
= & \delta \left( k - k' \right).
\end{align}
\end{widetext}

Eq. (\ref{vacuum_state_property}) can be proven from the definition of the Minkowski vacuum state $|0_M\rangle$
\begin{equation} \label{vacuum_state_H_M_in_tilde_H'_M_epsilon}
\hat{a}(k) |0_M\rangle = 0, \quad \forall k \in \mathbb{R},
\end{equation}
supplemented with Eqs. (\ref{Bogolyubov_transformation}):
\begin{align} \label{vacuum_state_property_proof}
& \int_{-\infty}^{+\infty}dK \left\lbrace\beta(k,K) \left[ \exp \left( \frac{\beta}{2} |K| \right)  \hat{b}_L(K) -  \hat{b}^\dagger_R(K) \right] \right. \nonumber \\
& + \left. \beta^*(k,K) \left[\exp \left( \frac{\beta}{2} |K| \right) \hat{b}_R(K) -  \hat{b}^\dagger_L(K) \right] \right\rbrace|0_M\rangle = 0.
\end{align}
Finally, Eq. (\ref{ordering_Rindler}) can be proven from Eq. (\ref{vacuum_state_property}):
\begin{align}
\hat{b}^\dagger_R(K) \hat{\rho}_0 = & \text{Tr}_L \left[ \hat{b}^\dagger_R(K) |0_M\rangle \langle 0_M| \right] \nonumber \\
= & \exp \left( \frac{1}{2} \beta |K| \right) \text{Tr}_L \left[ \hat{b}_L(K) |0_M\rangle \langle 0_M| \right] \nonumber \\
= & \exp \left( \frac{1}{2} \beta |K| \right) \text{Tr}_L \left[ |0_M\rangle \langle 0_M| \hat{b}_L(K) \right]\nonumber \\
= & e^{\beta |K|} \text{Tr}_L \left[ |0_M\rangle \langle 0_M| \hat{b}_R^\dagger(K) \right]\nonumber \\
= & e^{\beta |K|}  \hat{\rho}_0 \hat{b}^\dagger_R(K).
\end{align}

We rewrite Eq. (\ref{characteristic_function_step_1}) following step \ref{prescription_2} and step \ref{prescription_3}
\begin{align} \label{varrho_step_1}
|\psi\rangle \langle \psi | = & \sum_{n, n'} \psi[n] \psi^*[n'] \mathcal{K} \left( \prod_{k\in\mathbf{K}(n)} \hat{A}^\dagger(k) \right) \nonumber \\
& \times |0_M \rangle \langle 0_M | \mathcal{K} \left( \prod_{k'\in\mathbf{K}(n')} \hat{A}(k') \right),
\end{align}
where
\begin{subequations}
\begin{equation}
\hat{A}(k) = \hat{A}_\mathcal{L}(k) + \hat{A}_\mathcal{R}(k),
\end{equation}
\begin{align} \label{A_L}
\hat{A}_\mathcal{L}(k) = & \int_{-\infty}^{+\infty}dK \left[\exp \left( \frac{\beta}{2} |K| \right) \alpha(k,K)\hat{b}^\dagger_R(K) \right. \nonumber \\
& \left. - \exp \left(-\frac{\beta}{2} |K| \right) \beta^*(k,K) \hat{b}_R(K)\right],
\end{align}
\begin{equation} \label{A_R}
\hat{A}_\mathcal{R}(k) = \int_{-\infty}^{+\infty}dK \left[ \alpha^*(k,K) \hat{b}_R(K) - \beta(k,K) \hat{b}^\dagger_R(K) \right],
\end{equation}
\end{subequations}
and $\mathcal{K}$ defines a fixed ordering rule for $\hat{A}_\mathcal{L}(k)$, $\hat{A}_\mathcal{R}(k)$ and their adjoint operators depending on the arbitrary ordering for $a(k)$-operators in Eq. (\ref{characteristic_function_step_1_2}). For instance, we can order the $a(k)$-operators of Eq. (\ref{characteristic_function_step_1_2}) monotonically with respect to $k$ and obtain the following definition for $\mathcal{K}$:
\begin{subequations}
\begin{equation}
\mathcal{K}\left(\hat{A}_\mathcal{L}(k) \hat{A}_\mathcal{R}(k')\right) = \mathcal{K}\left(\hat{A}_\mathcal{R}(k') \hat{A}_\mathcal{L}(k)\right) = \hat{A}_\mathcal{L}(k) \hat{A}_\mathcal{R}(k'),
\end{equation}
\begin{equation}
\mathcal{K}\left(\hat{A}^\dagger_\mathcal{L}(k) \hat{A}^\dagger_\mathcal{R}(k')\right) = \mathcal{K}\left(\hat{A}^\dagger_\mathcal{R}(k') \hat{A}^\dagger_\mathcal{L}(k)\right) = \hat{A}^\dagger_\mathcal{R}(k) \hat{A}^\dagger_\mathcal{L}(k'),
\end{equation}
\begin{equation}
\mathcal{K}\left(\hat{A}_\mathcal{L}(k) \hat{A}_\mathcal{L}(k')\right) = \begin{cases}
\hat{A}_\mathcal{L}(k) \hat{A}_\mathcal{L}(k') & \text{if } k<k' \\
\hat{A}_\mathcal{L}(k') \hat{A}_\mathcal{L}(k) & \text{if } k>k' 
\end{cases},
\end{equation}
\begin{equation}
\mathcal{K}\left(\hat{A}_\mathcal{R}(k) \hat{A}_\mathcal{R}(k')\right) = \begin{cases}
\hat{A}_\mathcal{R}(k') \hat{A}_\mathcal{R}(k) & \text{if } k<k' \\
\hat{A}_\mathcal{R}(k) \hat{A}_\mathcal{R}(k') & \text{if } k>k' 
\end{cases},
\end{equation}
\begin{equation}
\mathcal{K}\left(\hat{A}^\dagger_\mathcal{L}(k) \hat{A}^\dagger_\mathcal{L}(k')\right) = \begin{cases}
\hat{A}^\dagger_\mathcal{L}(k') \hat{A}^\dagger_\mathcal{L}(k) & \text{if } k<k' \\
\hat{A}^\dagger_\mathcal{L}(k) \hat{A}^\dagger_\mathcal{L}(k') & \text{if } k>k' 
\end{cases},
\end{equation}
\begin{equation}
\mathcal{K}\left(\hat{A}^\dagger_\mathcal{R}(k) \hat{A}^\dagger_\mathcal{R}(k')\right) = \begin{cases}
\hat{A}^\dagger_\mathcal{R}(k) \hat{A}^\dagger_\mathcal{R}(k') & \text{if } k<k' \\
\hat{A}^\dagger_\mathcal{R}(k') \hat{A}^\dagger_\mathcal{R}(k) & \text{if } k>k' 
\end{cases}.
\end{equation}
\end{subequations}

The operators acting on the left and on the right of $|0_M \rangle \langle 0_M |$ in Eq. (\ref{varrho_step_1}) are combinations of chains of $\hat{b}_R$-operators. Each chain can be rewritten using the Wick theorem by considering $\hat{b}^\dagger_R(K)$ and $\hat{b}_R(K)$ as creation and annihilation operators for the normal ordering $\mathcal{N}$ -- i.e.
\begin{align}
& \mathcal{N} \left( \hat{b}^\dagger_R(K) \hat{b}_R(K') \right) = \mathcal{N} \left( \hat{b}_R(K') \hat{b}^\dagger_R(K)  \right) \nonumber \\
 = & \hat{b}^\dagger_R(K) \hat{b}_R(K')
\end{align}
-- and by defining $C_\mathcal{N}$ as a real function that can be evaluated on any chain of $\hat{b}_R$-operators and compute the sum of all the full contractions of such chain, with the following fundamental contractions:
\begin{subequations}
\begin{equation}
C_\mathcal{N} \left( \hat{b}_R(K) \hat{b}^\dagger_R(K') \right) = \left[ \hat{b}_R(K), \hat{b}^\dagger_R(K') \right] = \delta(K-K'),
\end{equation}
\begin{align}
& C_\mathcal{N} \left( \hat{b}^\dagger_R(K) \hat{b}_R(K') \right) = 0, & C_\mathcal{N} \left( \hat{b}^\dagger_R(K) \hat{b}^\dagger_R(K') \right) = 0, \\
& C_\mathcal{N} \left( \hat{b}_R(K) \hat{b}_R(K') \right) = 0.
\end{align}
\end{subequations}
The combination between the Wick theorem and the $\mathcal{K}$-ordering gives
\begin{widetext}
\begin{subequations} \label{normal_ordering_partial}
\begin{align}
\mathcal{K} \left( \prod_{k\in \mathbf{K}(n)} \hat{A}(k) \right)  = & \sum_{\mathbf{S}_0 \subseteq \mathbf{K}(n) } C_\mathcal{N} \left( \mathcal{K} \left( \prod_{k \in \mathbf{K}(n) \setminus \mathbf{S}_0} \hat{A}(k) \right) \right) \mathcal{N} \left( \mathcal{K} \left( \prod_{k_0 \in \mathbf{S}_0} \hat{A}(k_0) \right) \right) \nonumber \\
= & \sum_{\mathbf{S}_0 \subseteq \mathbf{K}(n) } C_\mathcal{K} \left( \prod_{k \in \mathbf{K}(n) \setminus \mathbf{S}_0}  \hat{A}(k) \right) \mathcal{N} \left(  \prod_{k_0 \in \mathbf{S}_0} \hat{A}(k_0) \right),
\end{align}
\begin{align}
\mathcal{K} \left( \prod_{k\in \mathbf{K}(n)} \hat{A}^\dagger(k) \right)  = & \sum_{\mathbf{S}_0 \subseteq \mathbf{K}(n) } C_\mathcal{N} \left( \mathcal{K} \left( \prod_{k \in \mathbf{K}(n) \setminus \mathbf{S}_0} \hat{A}^\dagger(k) \right) \right) \mathcal{N} \left( \mathcal{K} \left( \prod_{k_0 \in \mathbf{S}_0} \hat{A}^\dagger(k_0) \right) \right) \nonumber \\
= & \sum_{\mathbf{S}_0 \subseteq \mathbf{K}(n) } C_\mathcal{K} \left( \prod_{k \in \mathbf{K}(n) \setminus \mathbf{S}_0}  \hat{A}^\dagger(k) \right) \mathcal{N} \left(  \prod_{k_0 \in \mathbf{S}_0} \hat{A}^\dagger(k_0) \right),
\end{align}
\end{subequations}
with $C_\mathcal{K} = C_\mathcal{N} \circ \mathcal{K}$ as a new contraction such that
\begin{subequations}\label{fundamental_contractions_A}
\begin{align}
C_\mathcal{K} \left( \hat{A}(k) \hat{A}(k') \right)  = & C_\mathcal{N} \left( \mathcal{K} \left( \hat{A}(k) \hat{A}(k') \right) \right) \nonumber \\
= & \begin{cases}
 C_\mathcal{N} \left( \hat{A}_\mathcal{L}(k) \hat{A}_\mathcal{L}(k') + \hat{A}_\mathcal{L}(k) \hat{A}_\mathcal{R}(k') + \hat{A}_\mathcal{L}(k') \hat{A}_\mathcal{R}(k) + \hat{A}_\mathcal{R}(k') \hat{A}_\mathcal{R}(k) \right)  & \text{if } k<k' \\
C_\mathcal{N} \left( \hat{A}_\mathcal{L}(k') \hat{A}_\mathcal{L}(k) + \hat{A}_\mathcal{L}(k) \hat{A}_\mathcal{R}(k') + \hat{A}_\mathcal{L}(k') \hat{A}_\mathcal{R}(k) + \hat{A}_\mathcal{R}(k) \hat{A}_\mathcal{R}(k') \right)  & \text{if } k>k'
\end{cases} 
\end{align}
\begin{align}
C_\mathcal{K} \left( \hat{A}^\dagger(k) \hat{A}^\dagger(k') \right) = & C_\mathcal{N} \left( \mathcal{K} \left( \hat{A}^\dagger(k) \hat{A}^\dagger(k') \right) \right) \nonumber \\
= & \begin{cases}
 C_\mathcal{N} \left( \hat{A}^\dagger_\mathcal{L}(k') \hat{A}^\dagger_\mathcal{L}(k) + \hat{A}^\dagger_\mathcal{R}(k') \hat{A}^\dagger_\mathcal{L}(k) + \hat{A}^\dagger_\mathcal{R}(k) \hat{A}^\dagger_\mathcal{L}(k')  + \hat{A}^\dagger_\mathcal{R}(k) \hat{A}^\dagger_\mathcal{R}(k') \right)  & \text{if } k<k' \\
C_\mathcal{N} \left( \hat{A}^\dagger_\mathcal{L}(k)\hat{A}^\dagger_\mathcal{L}(k')  + \hat{A}^\dagger_\mathcal{R}(k') \hat{A}^\dagger_\mathcal{L}(k) + \hat{A}^\dagger_\mathcal{R}(k) \hat{A}^\dagger_\mathcal{L}(k') + \hat{A}^\dagger_\mathcal{R}(k') \hat{A}^\dagger_\mathcal{R}(k) \right)  & \text{if } k>k'
\end{cases}
\end{align}
\end{subequations}
Thanks to Eq. (\ref{alpha_to_beta}), Eq. (\ref{fundamental_contractions_A}) writes
\begin{equation}\label{fundamental_contractions_A_2}
C_\mathcal{K} \left( \hat{A}(k) \hat{A}(k') \right)  = C_\mathcal{K} \left( \hat{A}^\dagger(k) \hat{A}^\dagger(k') \right) = - \int_{-\infty}^{+\infty}dK 2 \sinh \left( \frac{\beta}{2} |K| \right) \left[ \beta^*(k,K) \beta(k',K) + \beta(k,K) \beta^*(k',K) \right].
\end{equation}
By combining Eq. (\ref{normal_ordering_partial}) with Eq. (\ref{fundamental_contractions_A_2}) we find a way to put Eq. (\ref{varrho_step_1}) in normal ordering at the left and right of $|0_M \rangle \langle 0_M |$:
\begin{align}
|\psi\rangle \langle \psi | = & \sum_{n, n'} \psi[n] \psi^*[n'] \sum_{\substack{\mathbf{S}_0 \subseteq \mathbf{K}(n) \\ \mathbf{S}'_0 \subseteq \mathbf{K}(n')  }} C_\mathcal{K} \left( \prod_{k \in \mathbf{K}(n) \setminus \mathbf{S}_0}   \hat{A}^\dagger(k) \right)  C_\mathcal{K} \left(  \prod_{k' \in \mathbf{K}(n') \setminus \mathbf{S}'_0} \hat{A}(k') \right)  \nonumber \\
& \times \mathcal{N} \left(  \prod_{k_0 \in \mathbf{S}_0} \hat{A}^\dagger(k_0) \right)  |0_M \rangle \langle 0_M | \mathcal{N} \left( \prod_{k'_0 \in \mathbf{S}'_0} \hat{A}(k'_0) \right)
\end{align}

Step \ref{prescription_4} gives
\begin{align} \label{varrho_step_4_1}
\hat{\rho} = & \text{Tr}_L |\psi\rangle \langle \psi | \nonumber \\
 = & \sum_{n, n'} \psi[n] \psi^*[n'] \sum_{\substack{\mathbf{S}_0 \subseteq \mathbf{K}(n) \\ \mathbf{S}'_0 \subseteq \mathbf{K}(n')  }} C_\mathcal{K} \left( \prod_{k \in \mathbf{K}(n) \setminus \mathbf{S}_0}   \hat{A}^\dagger(k) \right)  C_\mathcal{K} \left(  \prod_{k' \in \mathbf{K}(n') \setminus \mathbf{S}'_0} \hat{A}(k') \right) \mathcal{N} \left(  \prod_{k_0 \in \mathbf{S}_0} \hat{A}^\dagger(k_0) \right)  \hat{\rho}_0\mathcal{N} \left( \prod_{k'_0 \in \mathbf{S}'_0} \hat{A}(k'_0) \right)
\end{align}

We can explicitly compute the normal ordering of Eq. (\ref{varrho_step_4_1}) by giving a new decomposition for $\hat{A}(k)$:
\begin{subequations}
\begin{equation}
\hat{A}(k) = \hat{A}_+(k) + \hat{A}_-(k),
\end{equation}
\begin{equation} \label{A_+}
\hat{A}_+(k) =  \int_{-\infty}^{+\infty}dK \left[ \exp \left( \frac{\beta}{2} |K| \right) \alpha(k,K) - \beta(k,K) \right] \hat{b}^\dagger_R(K),
\end{equation}
\begin{equation} \label{A_-}
\hat{A}_-(k) =  \int_{-\infty}^{+\infty}dK \left[ - \exp \left(-\frac{\beta}{2} |K| \right) \beta^*(k,K) + \alpha^*(k,K) \right] \hat{b}_R(K)
\end{equation}
\end{subequations}
In this way Eq. (\ref{varrho_step_4_1}) reads
\begin{align} \label{varrho_step_4_2}
\hat{\rho} = & \sum_{n, n'} \psi[n] \psi^*[n'] \sum_{\substack{\mathbf{S}_0 \subseteq \mathbf{K}(n) \\ \mathbf{S}'_0 \subseteq \mathbf{K}(n')  }}  C_\mathcal{K} \left( \prod_{k \in \mathbf{K}(n) \setminus \mathbf{S}_0}  \hat{A}^\dagger(k) \right) C_\mathcal{K} \left(  \prod_{k' \in \mathbf{K}(n') \setminus \mathbf{S}'_0} \hat{A}(k') \right)  \nonumber \\
& \times  \sum_{\substack{\mathbf{S}_1 \subseteq \mathbf{S}_0 \\ \mathbf{S}'_1 \subseteq \mathbf{S}'_0}}  \prod_{k_0\in\mathbf{S}_0 \setminus \mathbf{S}_1} \hat{A}_-^\dagger(k_0)  \prod_{k_1\in\mathbf{S}_1} \hat{A}_+^\dagger(k_1)   \hat{\rho}_0  \prod_{k'_1\in\mathbf{S}'_1} \hat{A}_+(k'_1)  \prod_{k'_0\in\mathbf{S}'_0\setminus\mathbf{S}'_1} \hat{A}_-(k'_0).
\end{align}
Eqs. (\ref{A_+}, \ref{A_-}) can be put in the following form thanks to Eq. (\ref{alpha_to_beta})
\begin{align}
&\hat{A}_+(k) =  \int_{-\infty}^{+\infty}dK 2 \sinh \left(\frac{\beta}{2} |K| \right) \alpha(k,K)\hat{b}^\dagger_R(K), & \hat{A}_-(k) =  \int_{-\infty}^{+\infty}dK 2 \sinh \left(\frac{\beta}{2} |K| \right) \beta^*(k,K)\hat{b}_R(K)
\end{align}

By using Eq. (\ref{ordering_Rindler}) we can manipulate Eq. (\ref{varrho_step_4_2}) in the following way
\begin{align} \label{varrho_step_5_1}
\hat{\rho} = & \sum_{n, n'} \psi[n] \psi^*[n'] \sum_{\substack{\mathbf{S}_0 \subseteq \mathbf{K}(n) \\ \mathbf{S}'_0 \subseteq \mathbf{K}(n')  }}  C_\mathcal{K} \left( \prod_{k \in \mathbf{K}(n) \setminus \mathbf{S}_0}  \hat{A}^\dagger(k) \right) C_\mathcal{K} \left(  \prod_{k' \in \mathbf{K}(n') \setminus \mathbf{S}'_0} \hat{A}(k') \right)  \nonumber \\
& \times  \sum_{\substack{\mathbf{S}_1 \subseteq \mathbf{S}_0 \\ \mathbf{S}'_1 \subseteq \mathbf{S}'_0}}   \prod_{k_0\in\mathbf{S}_0 \setminus \mathbf{S}_1} \hat{A}_-^\dagger(k_0)  \prod_{k_1\in\mathbf{S}_1} \hat{A}_+^\dagger(k_1)  \prod_{k'_1\in\mathbf{S}'_1} \hat{B}_+(k'_1)  \hat{\rho}_0  \prod_{k'_0\in\mathbf{S}'_0\setminus\mathbf{S}'_1} \hat{A}_-(k'_0),
\end{align}
with
\begin{equation} \label{B_+}
\hat{B}_+(k) =  \int_{-\infty}^{+\infty}dK 2 \sinh \left(\frac{\beta}{2} |K| \right) e^{-\beta|K|} \alpha(k,K)\hat{b}^\dagger_R(K)
\end{equation}

By following step \ref{prescription_5}, we want to put the right side of Eq. (\ref{varrho_step_5_1}) in a normal order for the entire chain of $\hat{A}_\pm(k)$, $\hat{A^\dagger}_\pm(k)$, and $\hat{B}_+(k)$ operators. For this reason we will use again the Wick theorem for the $\hat{A}^\dagger_+ (k)$ and $\hat{B}_+ (k)$ operators:
\begin{align}
\hat{\rho} = & \sum_{n, n'} \psi[n] \psi^*[n'] \sum_{\substack{\mathbf{S}_0 \subseteq \mathbf{K}(n) \\ \mathbf{S}'_0 \subseteq \mathbf{K}(n')  }}  C_\mathcal{K} \left( \prod_{k \in \mathbf{K}(n) \setminus \mathbf{S}_0}  \hat{A}^\dagger(k) \right) C_\mathcal{K} \left(  \prod_{k' \in \mathbf{K}(n') \setminus \mathbf{S}'_0} \hat{A}(k') \right) \sum_{\substack{\mathbf{S}_1 \subseteq \mathbf{S}_0 \\ \mathbf{S}'_1 \subseteq \mathbf{S}'_0}}   \sum_{\substack{\mathbf{S}_2 \subseteq  \mathbf{S}_1 \\ \mathbf{S}'_2 \subseteq \mathbf{S}'_1 }}  C_\mathcal{N} \left(   \prod_{k_1 \in \mathbf{S}_1 \setminus \mathbf{S}_2} \hat{A}^\dagger_+(k_1) \right. \nonumber \\
& \left. \times  \prod_{k'_1 \in \mathbf{S}'_1 \setminus \mathbf{S}'_2} \hat{B}_+(k'_1) \right) \prod_{k_0\in\mathbf{S}_0 \setminus \mathbf{S}_1} \hat{A}_-^\dagger(k_0) \mathcal{N} \left( \prod_{k_2\in\mathbf{S}_2 } \hat{A}_+^\dagger(k_2)   \prod_{k'_2\in\mathbf{S}'_2} \hat{B}_+(k'_2) \right) \hat{\rho}_0  \prod_{k'_0\in\mathbf{S}'_0\setminus\mathbf{S}'_1} \hat{A}_-(k'_0) \nonumber \\
 = & \sum_{n, n'} \psi[n] \psi^*[n'] \sum_{\substack{\mathbf{S}_0 \subseteq \mathbf{K}(n) \\ \mathbf{S}'_0 \subseteq \mathbf{K}(n')  }}  C_\mathcal{K} \left( \prod_{k \in \mathbf{K}(n) \setminus \mathbf{S}_0}  \hat{A}^\dagger(k) \right) C_\mathcal{K} \left(  \prod_{k' \in \mathbf{K}(n') \setminus \mathbf{S}'_0} \hat{A}(k') \right) \sum_{\substack{\mathbf{S}_1 \subseteq \mathbf{S}_0 \\ \mathbf{S}'_1 \subseteq \mathbf{S}'_0}}   \sum_{\substack{\mathbf{S}_2 \subseteq  \mathbf{S}_1 \\ \mathbf{S}'_2 \subseteq \mathbf{S}'_1 }}  C_\mathcal{N} \left(   \prod_{k_1 \in \mathbf{S}_1 \setminus \mathbf{S}_2} \hat{A}^\dagger_+(k_1) \right. \nonumber \\
& \left. \times  \prod_{k'_1 \in \mathbf{S}'_1 \setminus \mathbf{S}'_2} \hat{B}_+(k'_1) \right) \prod_{k_0\in\mathbf{S}_0 \setminus \mathbf{S}_1} \hat{A}_-^\dagger(k_0) \prod_{k'_2\in\mathbf{S}'_2} \hat{B}_+(k'_2) \prod_{k_2\in\mathbf{S}_2 } \hat{A}_+^\dagger(k_2) \hat{\rho}_0  \prod_{k'_0\in\mathbf{S}'_0\setminus\mathbf{S}'_1} \hat{A}_-(k'_0) .
\end{align}
By defining $\mathbf{S}_3 = \mathbf{S}_0\setminus(\mathbf{S}_1\setminus\mathbf{S}_2)$ and $\mathbf{S}'_3 = \mathbf{S}'_0\setminus(\mathbf{S}'_1\setminus\mathbf{S}'_2)$, we obtain
\begin{align}
\hat{\rho} =  & \sum_{n, n'} \psi[n] \psi^*[n'] \sum_{\substack{\mathbf{S}_0 \subseteq \mathbf{K}(n) \\ \mathbf{S}'_0 \subseteq \mathbf{K}(n')  }}  C_\mathcal{K} \left( \prod_{k \in \mathbf{K}(n) \setminus \mathbf{S}_0}  \hat{A}^\dagger(k) \right) C_\mathcal{K} \left(  \prod_{k' \in \mathbf{K}(n') \setminus \mathbf{S}'_0} \hat{A}(k') \right) \sum_{\substack{\mathbf{S}_3 \subseteq \mathbf{S}_0 \\ \mathbf{S}'_3 \subseteq \mathbf{S}'_0 }}   C_\mathcal{N} \left(   \prod_{k_1 \in \mathbf{S}_0 \setminus \mathbf{S}_3} \hat{A}^\dagger_+(k_1) \right. \nonumber \\
& \left. \times  \prod_{k'_1 \in \mathbf{S}'_0 \setminus \mathbf{S}'_3} \hat{B}_+(k'_1) \right) \sum_{\substack{\mathbf{S}_2 \subseteq \mathbf{S}_3 \\ \mathbf{S}'_2 \subseteq \mathbf{S}'_3 }}  \prod_{k_0\in\mathbf{S}_3\setminus\mathbf{S}_2} \hat{A}_-^\dagger(k_0)  \prod_{k'_2\in\mathbf{S}'_2} \hat{B}_+(k'_2) \prod_{k_2\in\mathbf{S}_2} \hat{A}_+^\dagger(k_2)  \hat{\rho}_0  \prod_{k'_0\in\mathbf{S}'_3\setminus\mathbf{S}'_2} \hat{A}_-(k'_0).
\end{align}
Finally, by defining $\mathbf{S}_4 = \mathbf{S}_0 \setminus \mathbf{S}_3$ and $\mathbf{S}'_4 = \mathbf{S}'_0 \setminus \mathbf{S}'_3$, we obtain
\begin{align}\label{varrho_step_5_2}
\hat{\rho} =  & \sum_{n, n'} \psi[n] \psi^*[n'] \sum_{\substack{\mathbf{S}_3 \subseteq \mathbf{K}(n) \\ \mathbf{S}'_3 \subseteq \mathbf{K}(n')  }}  \sum_{\substack{\mathbf{S}_4 \subseteq  \mathbf{K}(n)  \setminus \mathbf{S}_3 \\ \mathbf{S}'_4 \subseteq  \mathbf{K}(n') \setminus \mathbf{S}'_3  }}  C_\mathcal{K} \left( \prod_{k \in \mathbf{K}(n) \setminus (\mathbf{S}_3 \cup \mathbf{S}_4)}  \hat{A}^\dagger(k) \right) C_\mathcal{K} \left(  \prod_{k' \in \mathbf{K}(n') \setminus (\mathbf{S}'_3 \cup \mathbf{S}'_4)} \hat{A}(k') \right)  \nonumber \\
&  \times C_\mathcal{N} \left(   \prod_{k_4 \in  \mathbf{S}_4} \hat{A}^\dagger_+(k_4)  \prod_{k'_4 \in \mathbf{S}'_4} \hat{B}_+(k'_4) \right) \sum_{\substack{\mathbf{S}_2 \subseteq \mathbf{S}_3 \\ \mathbf{S}'_2 \subseteq \mathbf{S}'_3 }}  \prod_{k_0\in\mathbf{S}_3\setminus\mathbf{S}_2} \hat{A}_-^\dagger(k_0)  \prod_{k'_2\in\mathbf{S}'_2} \hat{B}_+(k'_2) \prod_{k_2\in\mathbf{S}_2} \hat{A}_+^\dagger(k_2)  \hat{\rho}_0  \prod_{k'_0\in\mathbf{S}'_3\setminus\mathbf{S}'_2} \hat{A}_-(k'_0).
\end{align}

The full contractions appearing in Eq. (\ref{varrho_step_5_2}) can be manipulated in a combinatoric way by knowing that
\begin{subequations}
\begin{align}
& C_\mathcal{N} \left( \hat{A}^\dagger_+(k) \hat{A}^\dagger_+(k') \right) = 0, & C_\mathcal{N}  \left( \hat{B}_+(k) \hat{B}_+(k') \right) = 0,
\end{align}
\begin{equation}
C_\mathcal{N}  \left( \hat{A}^\dagger_+(k) \hat{B}_+(k') \right) =  \int_{-\infty}^{+\infty}dK 2 \sinh \left(\frac{\beta}{2} |K| \right) 2 \sinh \left(\frac{\beta}{2} |K| \right) e^{-\beta|K|}  \alpha^*(k,K) \alpha(k',K) 
\end{equation}
\end{subequations}
This allows us to put the right side part of Eq. (\ref{varrho_step_5_2}) in a more compact way, by defining the following new contraction $C$:
\begin{subequations}
\begin{equation}
C \left( \hat{A}(k) \hat{A}(k') \right)  = C_\mathcal{K} \left( \hat{A}(k) \hat{A}(k') \right) = - \int_{-\infty}^{+\infty}dK 2 \sinh \left( \frac{\beta}{2} |K| \right) \left[ \beta^*(k,K) \beta(k',K) + \beta(k,K) \beta^*(k',K) \right],
\end{equation}
\begin{equation}
C \left( \hat{A}^\dagger(k) \hat{A}^\dagger(k') \right) = C_\mathcal{K} \left( \hat{A}^\dagger(k) \hat{A}^\dagger(k') \right) = - \int_{-\infty}^{+\infty}dK 2 \sinh \left( \frac{\beta}{2} |K| \right) \left[ \beta^*(k,K) \beta(k',K) + \beta(k,K) \beta^*(k',K) \right],
\end{equation}
\begin{align}
C \left( \hat{A}^\dagger(k) \hat{A}(k') \right) = C_\mathcal{N}  \left( \hat{A}^\dagger_+(k) \hat{B}_+(k') \right) =  \int_{-\infty}^{+\infty}dK 2 \sinh \left(\frac{\beta}{2} |K| \right) 2 \sinh \left(\frac{\beta}{2} |K| \right) e^{-\beta|K|}  \alpha^*(k,K) \alpha(k',K).
\end{align}
\end{subequations}
In this way Eq. (\ref{varrho_step_5_2}) now writes
\begin{align}\label{varrho_step_5_3}
\hat{\rho} =  & \sum_{n, n'} \psi[n] \psi^*[n'] \sum_{\substack{\mathbf{S}_3 \subseteq \mathbf{K}(n) \\ \mathbf{S}'_3 \subseteq \mathbf{K}(n')  }}  C \left( \prod_{k \in  \mathbf{K}(n) \setminus \mathbf{S}_3}  \hat{A}^\dagger(k) \prod_{k' \in \mathbf{K}(n') \setminus \mathbf{S}'_3} \hat{A}(k') \right)  \nonumber \\
& \times \sum_{\substack{\mathbf{S}_2 \subseteq \mathbf{S}_3 \\ \mathbf{S}'_2 \subseteq \mathbf{S}'_3 }}   \prod_{k_0\in\mathbf{S}_3\setminus\mathbf{S}_2} \hat{A}_-^\dagger(k_0)  \prod_{k'_2\in\mathbf{S}'_2} \hat{B}_+(k'_2) \prod_{k_2\in\mathbf{S}_2} \hat{A}_+^\dagger(k_2)  \hat{\rho}_0  \prod_{k'_0\in\mathbf{S}'_3\setminus\mathbf{S}'_2} \hat{A}_-(k'_0).
\end{align}
It is possible to notice that the $C$-contraction of chains of $\hat{A}(k)$- and $\hat{A}^\dagger(k)$-operators does not depend of their order within such chains, therefore we write $C$ as a function of sets of momenta:
\begin{equation}
C \left( \prod_{k \in \mathbf{U}}  \hat{A}^\dagger(k) \prod_{k' \in \mathbf{U}'} \hat{A}(k') \right) =  C (\mathbf{U},\mathbf{U}').
\end{equation}
The same convention will be used for any other ordering-invariant contraction. Thanks to the definition of $C (\mathbf{U},\mathbf{U}')$, Eq. (\ref{varrho_step_5_3}) writes
\begin{align}
\hat{\rho} = & \sum_{n, n'} \psi[n] \psi^*[n'] \sum_{\substack{\mathbf{S}_3 \subseteq \mathbf{K}(n) \\ \mathbf{S}'_3 \subseteq \mathbf{K}(n')  }} C (\mathbf{K}(n) \setminus \mathbf{S}_3,\mathbf{K}(n') \setminus \mathbf{S}'_3) \nonumber \\
& \times \sum_{\substack{\mathbf{S}_2 \subseteq \mathbf{S}_3 \\ \mathbf{S}'_2 \subseteq \mathbf{S}'_3 }}  \prod_{k_0\in\mathbf{S}_3\setminus\mathbf{S}_2} \hat{A}_-^\dagger(k_0)  \prod_{k'_2\in\mathbf{S}'_2} \hat{B}_+(k'_2) \prod_{k_2\in\mathbf{S}_2} \hat{A}_+^\dagger(k_2)  \hat{\rho}_0  \prod_{k'_0\in\mathbf{S}'_3\setminus\mathbf{S}'_2} \hat{A}_-(k'_0).
\end{align}

By using again Eq. (\ref{ordering_Rindler}) on $\hat{A}_+^\dagger(k)$-operators we conclude step \ref{prescription_5}
\begin{align}
\hat{\rho} = & \sum_{n, n'} \psi[n] \psi^*[n'] \sum_{\substack{\mathbf{S}_3 \subseteq \mathbf{K}(n) \\ \mathbf{S}'_3 \subseteq \mathbf{K}(n')  }}  C (\mathbf{K}(n) \setminus \mathbf{S}_3,\mathbf{K}(n') \setminus \mathbf{S}'_3) \nonumber \\
& \times \sum_{\substack{\mathbf{S}_2 \subseteq \mathbf{S}_3 \\ \mathbf{S}'_2 \subseteq \mathbf{S}'_3 }}  \prod_{k_0\in\mathbf{S}_3\setminus\mathbf{S}_2} \hat{A}_-^\dagger(k_0)  \prod_{k'_2\in\mathbf{S}'_2} \hat{B}_+(k'_2)  \hat{\rho}_0 \prod_{k_2\in\mathbf{S}_2} \hat{B}_+^\dagger(k_2)  \prod_{k'_0\in\mathbf{S}'_3\setminus\mathbf{S}'_2} \hat{A}_-(k'_0).
\end{align}

Step \ref{prescription_6} gives
\begin{align}\label{varrho_step_6}
\chi^{(-1)}[\xi, \xi^*] = & \text{Tr} \left( \hat{\rho} \hat{D}_{-1}[\xi, \xi^*] \right) \nonumber \\
 =  & \sum_{n, n'} \psi[n] \psi^*[n'] \sum_{\substack{\mathbf{S}_3 \subseteq \mathbf{K}(n) \\ \mathbf{S}'_3 \subseteq \mathbf{K}(n')  }} C (\mathbf{K}(n) \setminus \mathbf{S}_3,\mathbf{K}(n') \setminus \mathbf{S}'_3)   \nonumber \\
 & \times \sum_{\substack{\mathbf{S}_2 \subseteq \mathbf{S}_3 \\ \mathbf{S}'_2 \subseteq \mathbf{S}'_3 }}  \text{Tr} \left(  \prod_{k_0\in\mathbf{S}_3\setminus\mathbf{S}_2} \hat{A}_-^\dagger(k_0)  \prod_{k'_2\in\mathbf{S}'_2} \hat{B}_+(k'_2)  \hat{\rho}_0 \prod_{k_2\in\mathbf{S}_2} \hat{B}_+^\dagger(k_2)  \prod_{k'_0\in\mathbf{S}'_3\setminus\mathbf{S}'_2} \hat{A}_-(k'_0) \hat{D}_{-1}[\xi, \xi^*] \right) \nonumber \\
 =  & \sum_{n, n'} \psi[n] \psi^*[n'] \sum_{\substack{\mathbf{S}_3 \subseteq \mathbf{K}(n) \\ \mathbf{S}'_3 \subseteq \mathbf{K}(n')  }}  C (\mathbf{K}(n) \setminus \mathbf{S}_3,\mathbf{K}(n') \setminus \mathbf{S}'_3)  \nonumber \\
 & \times \sum_{\substack{\mathbf{S}_2 \subseteq \mathbf{S}_3 \\ \mathbf{S}'_2 \subseteq \mathbf{S}'_3 }}  \text{Tr} \left(  \hat{\rho}_0 \prod_{k_2\in\mathbf{S}_2} \hat{B}_+^\dagger(k_2)  \prod_{k'_0\in\mathbf{S}'_3\setminus\mathbf{S}'_2} \hat{A}_-(k'_0) \hat{D}_{-1}[\xi, \xi^*] \prod_{k_0\in\mathbf{S}_3\setminus\mathbf{S}_2} \hat{A}_-^\dagger(k_0)  \prod_{k'_2\in\mathbf{S}'_2} \hat{B}_+(k'_2) \right)
\end{align}

As prescribed by step \ref{prescription_7}, we manipulate Eq. (\ref{varrho_step_6}) by using Eqs. (\ref{multiple_Xi_derivatives_-1}):
\begin{align}\label{varrho_step_7_2}
\chi^{(-1)}[\xi, \xi^*] =  & \sum_{n, n'} \psi[n] \psi^*[n'] \sum_{\substack{\mathbf{S}_3 \subseteq \mathbf{K}(n) \\ \mathbf{S}'_3 \subseteq \mathbf{K}(n')  }}  C (\mathbf{K}(n) \setminus \mathbf{S}_3,\mathbf{K}(n') \setminus \mathbf{S}'_3)   \nonumber \\
 & \times \sum_{\substack{\mathbf{S}_2 \subseteq \mathbf{S}_3 \\ \mathbf{S}'_2 \subseteq \mathbf{S}'_3 }} \prod_{k_0\in\mathbf{S}_3\setminus\mathbf{S}_2} \overrightarrow{\Delta}_A(k_0)  \prod_{k'_2\in\mathbf{S}'_2} \overrightarrow{\Delta}_B(k'_2) \prod_{k_2\in\mathbf{S}_2} \left[ - \overrightarrow{\Delta}_B^*(k_2) \right]  \prod_{k'_0\in\mathbf{S}'_3\setminus\mathbf{S}'_2} \left[ - \overrightarrow{\Delta}_A^*(k'_0) \right] \text{Tr} \left(  \hat{\rho}_0 \hat{D}_{-1}[\xi, \xi^*] \right),
\end{align}
with
\begin{align}
& \overrightarrow{\Delta}_A(k) = \int_{-\infty}^{+\infty}dK 2 \sinh \left(\frac{\beta}{2} |K| \right) \beta(k,K)\frac{\delta}{\delta \xi (K)}, & \overrightarrow{\Delta}_B(k) = \int_{-\infty}^{+\infty}dK 2 \sinh \left(\frac{\beta}{2} |K| \right) e^{-\beta|K|} \alpha(k,K)\frac{\delta}{\delta \xi (K)}
\end{align}
as derivatives acting on their right.

By using Eq. (\ref{thermal_characteristic_fucntion}), we obtain
\begin{align}\label{varrho_step_7_3}
\chi^{(-1)}[\xi, \xi^*] =  & \sum_{n, n'} \psi[n] \psi^*[n'] \sum_{\substack{\mathbf{S}_3 \subseteq \mathbf{K}(n) \\ \mathbf{S}'_3 \subseteq \mathbf{K}(n')  }} C (\mathbf{K}(n) \setminus \mathbf{S}_3,\mathbf{K}(n') \setminus \mathbf{S}'_3)  \nonumber \\
 & \times \sum_{\substack{\mathbf{S}_2 \subseteq \mathbf{S}_3 \\ \mathbf{S}'_2 \subseteq \mathbf{S}'_3 }} \prod_{k_0\in\mathbf{S}_3\setminus\mathbf{S}_2} \overrightarrow{\Delta}_A(k_0)  \prod_{k'_2\in\mathbf{S}'_2} \overrightarrow{\Delta}_B(k'_2) \prod_{k_2\in\mathbf{S}_2} \left[ - \overrightarrow{\Delta}_B^*(k_2) \right] \prod_{k'_0\in\mathbf{S}'_3\setminus\mathbf{S}'_2} \left[ - \overrightarrow{\Delta}_A^*(k'_0) \right] \chi^{(-1)}_0[\xi,\xi^*] \nonumber \\
 =  & \sum_{n, n'} \psi[n] \psi^*[n'] \sum_{\substack{\mathbf{S}_3 \subseteq \mathbf{K}(n) \\ \mathbf{S}'_3 \subseteq \mathbf{K}(n')  }}  C (\mathbf{K}(n) \setminus \mathbf{S}_3,\mathbf{K}(n') \setminus \mathbf{S}'_3) \nonumber \\
 & \times \sum_{\substack{\mathbf{S}_2 \subseteq \mathbf{S}_3 \\ \mathbf{S}'_2 \subseteq \mathbf{S}'_3 }} \prod_{k_0\in\mathbf{S}_3\setminus\mathbf{S}_2} \overrightarrow{\Delta}_A(k_0)  \prod_{k'_2\in\mathbf{S}'_2} \overrightarrow{\Delta}_B(k'_2) \prod_{k_2\in\mathbf{S}_2} L_B (k_2) [\xi] \prod_{k'_0\in\mathbf{S}'_3\setminus\mathbf{S}'_2}  L_A (k'_0) [\xi]  \chi^{(-1)}_0[\xi,\xi^*],
\end{align}
with
\begin{subequations} \label{L_AB}
\begin{align}
L_B (k) [\xi] = & \int_{-\infty}^{+\infty}dK 2 \sinh \left(\frac{\beta}{2} |K| \right) e^{-\beta|K|} \alpha^*(k,K) (n_0+1) \xi(K) ,
\end{align}
\begin{align}
L_A (k) [\xi]  = & \int_{-\infty}^{+\infty}dK 2 \sinh \left(\frac{\beta}{2} |K| \right) \beta^*(k,K) (n_0+1) \xi(K) .
\end{align}
\end{subequations}
Eq. (\ref{L_AB}) can be computed thanks to the help of Eq. (\ref{alpha_to_beta}):
\begin{subequations}
\begin{align}
L_B (k) [\xi]  = & \int_{-\infty}^{+\infty}dK 2 \sinh \left(\frac{\beta}{2} |K| \right) \exp \left(-\frac{\beta}{2} |K| \right) \beta^*(k,K) (n_0+1) \xi(K) \nonumber \\
 = & \int_{-\infty}^{+\infty}dK \left( 1-e^{-\beta |K|} \right)  \left( \frac{1}{e^{\beta|K|}-1} +1 \right) \beta^*(k,K) \xi(K) \nonumber \\
 = & \int_{-\infty}^{+\infty}dK \frac{e^{\beta |K|}-1}{e^{\beta |K|}}  \frac{e^{\beta|K|}}{e^{\beta|K|}-1} \beta^*(k,K) \xi(K) \nonumber \\
= & \int_{-\infty}^{+\infty}dK \beta^*(k,K) \xi(K),
\end{align}
\begin{align}
L_A (k) [\xi] = & \int_{-\infty}^{+\infty}dK 2 \sinh \left(\frac{\beta}{2} |K| \right) \exp \left(-\frac{\beta}{2} |K| \right) \alpha^*(k,K) (n_0+1) \xi(K) \nonumber \\
 = & \int_{-\infty}^{+\infty}dK \left( 1-e^{-\beta |K|} \right) \left( \frac{1}{e^{\beta|K|}-1} +1 \right) \alpha^*(k,K) \xi(K)\nonumber \\
 = & \int_{-\infty}^{+\infty}dK \frac{e^{\beta |K|}-1}{e^{\beta |K|}}  \frac{e^{\beta|K|}}{e^{\beta|K|}-1} \alpha^*(k,K) \xi(K)\nonumber \\
 = & \int_{-\infty}^{+\infty}dK \alpha^*(k,K) \xi(K).
\end{align}
\end{subequations}

The derivatives $\overrightarrow{\Delta}_{A,B}(k) $ now have to be evaluated on both  $L_{A,B} (k) [\xi]$ and  $\chi^{(-1)}_0[\xi,\xi^*]$. In order to simplify the calculation, we define $\overleftarrow{\Delta}_{A,B}(k) $ as derivatives identical to $\overrightarrow{\Delta}_{A,B}(k) $ but acting on their left. Moreover, we define $\overleftrightarrow{\Delta}_{A,B}(k) = \overleftarrow{\Delta}_{A,B}(k) + \overrightarrow{\Delta}_{A,B}(k) $. In this way, Eq. (\ref{varrho_step_7_3}) can be put in a more compact form:
\begin{align}\label{varrho_step_7_4}
\chi^{(-1)}[\xi, \xi^*] =  & \sum_{n, n'} \psi[n] \psi^*[n'] \sum_{\substack{\mathbf{S}_3 \subseteq \mathbf{K}(n) \\ \mathbf{S}'_3 \subseteq \mathbf{K}(n')  }}  C (\mathbf{K}(n) \setminus \mathbf{S}_3,\mathbf{K}(n') \setminus \mathbf{S}'_3) \nonumber \\
 & \times \prod_{k_3 \in\mathbf{S}_3} \left\lbrace \overleftrightarrow{\Delta}_A(k_3) + L_B (k_3) [\xi] \right\rbrace \prod_{k'_3\in\mathbf{S}'_3} \left\lbrace \overleftrightarrow{\Delta}_B(k'_3) + L_A (k'_3) [\xi] \right\rbrace \chi^{(-1)}_0[\xi,\xi^*],
\end{align}
A further simplification can be made by defining $C_\Delta(\mathbf{U} , \mathbf{U}')$ as a contraction with the following fundamental contractions
\begin{subequations} \label{C_Delta}
\begin{equation}
C_\Delta(\{k,k' \} ,\varnothing) =  \overrightarrow{\Delta}_A(k)  L_B (k') [\xi] +  \overrightarrow{\Delta}_A(k')  L_B (k) [\xi],
\end{equation}
\begin{equation}
C_\Delta(\varnothing,\{k,k' \})  =   \overrightarrow{\Delta}_B(k)  L_A (k') [\xi] +  \overrightarrow{\Delta}_B(k')  L_A (k) [\xi],
\end{equation}
\begin{equation}
C_\Delta(\{k \},\{k' \})  =  \overrightarrow{\Delta}_A(k)  L_A (k') [\xi] + \overrightarrow{\Delta}_B(k')  L_B (k) [\xi]
\end{equation}
\end{subequations}
and by using the following identities, with, again, the help of Eq. (\ref{alpha_to_beta})
\begin{subequations}
\begin{align}
& \left\lbrace \overrightarrow{\Delta}_A(k) + L_B (k) [\xi] \right\rbrace  \chi^{(-1)}_0[\xi,\xi^*] \nonumber \\
 = &\int_{-\infty}^{+\infty}dK  \left[ 2 \sinh \left(\frac{\beta}{2} |K| \right) \beta(k,K)\frac{\delta}{\delta \xi (K)} + \beta^*(k,K) \xi(K) \right]   \chi^{(-1)}_0[\xi,\xi^*] \nonumber \\
= &\int_{-\infty}^{+\infty}dK  \left\lbrace -2 \sinh \left(\frac{\beta}{2} |K| \right) \beta(k,K)[n_0(K)+1]\xi ^*(K) + \beta^*(k,K) \xi(K) \right\rbrace   \chi^{(-1)}_0[\xi,\xi^*] \nonumber \\
= &\int_{-\infty}^{+\infty}dK  \left[ - \left( e^{\beta|K|} - 1 \right) \exp \left(-\frac{\beta}{2} |K| \right) \beta(k,K) \left(\frac{1}{e^{\beta|K|}-1} +1\right)\xi ^*(K) + \beta^*(k,K) \xi(K) \right]   \chi^{(-1)}_0[\xi,\xi^*] \nonumber \\
= &\int_{-\infty}^{+\infty}dK  \left[ -  \exp \left(\frac{\beta}{2} |K| \right) \beta(k,K) \xi ^*(K) + \beta^*(k,K) \xi(K) \right]   \chi^{(-1)}_0[\xi,\xi^*] \nonumber \\
= &\int_{-\infty}^{+\infty}dK  \left[ -  \alpha(k,K) \xi ^*(K) + \beta^*(k,K) \xi(K) \right]   \chi^{(-1)}_0[\xi,\xi^*] \nonumber \\
= & \{ - L (k) [\xi,\xi^*] \}^* \chi^{(-1)}_0[\xi,\xi^*],
\end{align}
\begin{align}
& \left\lbrace \overrightarrow{\Delta}_B(k) + L_A (k) [\xi] \right\rbrace \chi^{(-1)}_0[\xi,\xi^*] \nonumber \\
= & \int_{-\infty}^{+\infty}dK \left[ 2 \sinh \left(\frac{\beta}{2} |K| \right) e^{-\beta|K|} \alpha(k,K)\frac{\delta}{\delta \xi (K)} + \alpha^*(k,K) \xi(K) \right] \chi^{(-1)}_0[\xi,\xi^*] \nonumber \\
= & \int_{-\infty}^{+\infty}dK \left\lbrace -2 \sinh \left(\frac{\beta}{2} |K| \right) e^{-\beta|K|} \alpha(k,K)[n_0(K)+1]\xi^* (K) +  \alpha^*(k,K) \xi(K) \right\rbrace \chi^{(-1)}_0[\xi,\xi^*]\nonumber \\
= &\int_{-\infty}^{+\infty}dK  \left[ - \left( e^{\beta|K|} - 1 \right) \exp \left(-\frac{\beta}{2} |K| \right) e^{-\beta|K|} \alpha(k,K) \left(\frac{1}{e^{\beta|K|}-1} +1\right)\xi ^*(K) + \alpha^*(k,K) \xi(K) \right] \chi^{(-1)}_0[\xi,\xi^*] \nonumber \\
= &\int_{-\infty}^{+\infty}dK  \left[ - \exp \left(-\frac{\beta}{2} |K| \right) \alpha(k,K) \xi ^*(K) + \alpha^*(k,K) \xi(K) \right]   \chi^{(-1)}_0[\xi,\xi^*] \nonumber \\
= &\int_{-\infty}^{+\infty}dK  \left[ - \beta(k,K) \xi ^*(K) + \alpha^*(k,K) \xi(K) \right]   \chi^{(-1)}_0[\xi,\xi^*] \nonumber \\
= &  L (k) [\xi,\xi^*] \chi^{(-1)}_0[\xi,\xi^*].
\end{align}
\end{subequations}
In this way, Eq. (\ref{varrho_step_7_4}) can be computed in the following way:
\begin{align}\label{varrho_step_7_5}
\chi^{(-1)}[\xi, \xi^*] =  & \sum_{n, n'} \psi[n] \psi^*[n'] \sum_{\substack{\mathbf{S}_3 \subseteq \mathbf{K}(n) \\ \mathbf{S}'_3 \subseteq \mathbf{K}(n')  }}  C (\mathbf{K}(n) \setminus \mathbf{S}_3,\mathbf{K}(n') \setminus \mathbf{S}'_3) \sum_{\substack{\mathbf{S} \subseteq \mathbf{S}_3 \\ \mathbf{S}' \subseteq \mathbf{S}'_3  }} C_\Delta (\mathbf{S}_3 \setminus \mathbf{S},\mathbf{S}'_3 \setminus \mathbf{S}')  \nonumber \\
 & \times \prod_{k\in\mathbf{S}} \{ - L (k) [\xi,\xi^*] \}^* \prod_{k'\in\mathbf{S}'} L (k') [\xi,\xi^*]  \chi^{(-1)}_0[\xi,\xi^*],
\end{align}

The foundamental contractions of $C_\Delta(\mathbf{U} , \mathbf{U}')$ defined in Eqs. (\ref{C_Delta}) can be computed with the help of Eq. (\ref{alpha_to_beta})
\begin{subequations}
\begin{align}
C_\Delta(\{k,k' \} ,\varnothing) = &  \int_{-\infty}^{+\infty}dK 2 \sinh \left(\frac{\beta}{2} |K| \right) [\beta(k,K) \beta^*(k',K)+\beta^*(k,K) \beta(k',K)] \nonumber \\
 = &  - C(\{k,k' \} ,\varnothing)
\end{align}
\begin{align}
C_\Delta(\varnothing,\{k,k' \})  = & \int_{-\infty}^{+\infty}dK 2 \sinh \left(\frac{\beta}{2} |K| \right) e^{-\beta|K|} [\alpha(k,K) \alpha^*(k',K)+\alpha^*(k,K) \alpha(k',K)]\nonumber \\
= & \int_{-\infty}^{+\infty}dK 2 \sinh \left(\frac{\beta}{2} |K| \right) [\beta(k,K) \beta^*(k',K)+\beta^*(k,K) \beta(k',K)] \nonumber \\
 = &  - C(\varnothing,\{k,k' \})
\end{align}
\begin{align}
C_\Delta(\{k \},\{k' \})  = & \int_{-\infty}^{+\infty}dK 2 \sinh \left(\frac{\beta}{2} |K| \right)[ \beta(k,K) \alpha^*(k',K) +   e^{-\beta|K|} \beta^*(k,K) \alpha(k',K) ]
\end{align}
\end{subequations}

The contractions $C(\mathbf{U} , \mathbf{U}')$ and $C_\Delta(\mathbf{U} , \mathbf{U}')$ appearing in Eq. (\ref{varrho_step_7_5}) can be combined in a single contraction $c(\mathbf{U} , \mathbf{U}') = C(\mathbf{U} , \mathbf{U}') + C_\Delta(\mathbf{U} , \mathbf{U}')$ which has the following fundamental contractions
\begin{subequations}\label{c}
\begin{align}\label{c_a}
& c(\{k,k' \} ,\varnothing) =  C(\{k,k' \} ,\varnothing) + C_\Delta(\{k,k' \} ,\varnothing)  =  0, & c(\varnothing,\{k,k' \})  =  C(\varnothing,\{k,k' \}) + C_\Delta(\varnothing,\{k,k' \}) = 0,
\end{align}
\begin{align}
c(\{k \},\{k' \})  = &  C(\{k \},\{k' \}) + C_\Delta(\{k \},\{k' \}) \nonumber \\
 = &   \int_{-\infty}^{+\infty}dK 2 \sinh \left(\frac{\beta}{2} |K| \right)  \left[  2 \sinh \left(\frac{\beta}{2} |K| \right) e^{-\beta|K|}  \alpha^*(k,K) \alpha(k',K)\right. \nonumber \\
 & \left. +  \beta(k,K) \alpha^*(k',K) +   e^{-\beta|K|} \beta^*(k,K) \alpha(k',K)  \right]\nonumber \\
 = &   \int_{-\infty}^{+\infty}dK 2 \sinh \left(\frac{\beta}{2} |K| \right)  \left[ \exp \left(- \frac{\beta}{2} |K| \right)  \alpha^*(k,K) \alpha(k',K) - \exp \left(- \frac{3}{2} \beta |K| \right)  \alpha^*(k,K) \alpha(k',K)\right. \nonumber \\
 & \left. +  \beta(k,K) \alpha^*(k',K) +   e^{-\beta|K|} \beta^*(k,K) \alpha(k',K)  \right].
\end{align}
\end{subequations}
The last contraction can be computed through Eqs. (\ref{alpha_to_beta}, \ref{identity_K}):
\begin{equation}\label{c_b}
c(\{k \},\{k' \})  = \int_{-\infty}^{+\infty}dK 2 \sinh \left(\frac{\beta}{2} |K| \right)  [\alpha^*(k,K) \beta(k',K) +  \beta(k,K) \alpha^*(k',K) ] = \delta(k-k').
\end{equation}
The coefficients $c(\mathbf{U} , \mathbf{U}')$ of Eq. (\ref{c_combinatory}) are identical to the contractions $c(\mathbf{U} , \mathbf{U}')$ defined by Eqs. (\ref{c_a}, \ref{c_b}). Thanks to the definition of $c(\mathbf{U} , \mathbf{U}') = C(\mathbf{U} , \mathbf{U}') + C_\Delta(\mathbf{U} , \mathbf{U}')$, Eq. (\ref{varrho_step_7_5}) writes
\begin{equation}\label{varrho_step_7_6}
\chi^{(-1)}[\xi, \xi^*] =  \sum_{n, n'} \psi[n] \psi^*[n'] \sum_{\substack{\mathbf{S} \subseteq \mathbf{K}(n) \\ \mathbf{S}' \subseteq \mathbf{K}(n')  }}  c (\mathbf{K}(n) \setminus \mathbf{S},\mathbf{K}(n') \setminus \mathbf{S}') \prod_{k\in\mathbf{S}} \{ - L (k) [\xi,\xi^*] \}^* \prod_{k'\in\mathbf{S}'} L (k') [\xi,\xi^*]  \chi^{(-1)}_0[\xi,\xi^*].
\end{equation}
In this way we have concluded step \ref{prescription_7}.

Finally, with step \ref{prescription_8}, we obtain Eq.(\ref{chi_final}).

\section{A proof for Eq. (\ref{chi_M_final})}\label{A_proof_for_2}

Eq. (\ref{A_proof_for_2}) with $p=-1$ can be proved in the following way:
\begin{align}
\chi^{(-1)}_M[\xi_M, \xi^*_M] = & \text{Tr} \left( |\psi\rangle \langle \psi | \exp \left( - \int_{-\infty}^{+\infty} dk  \xi_M^*(k) \hat{a}(k) \right) \exp \left( \int_{-\infty}^{+\infty} dk \xi_M(k) \hat{a}^\dagger(k)  \right) \right) \nonumber \\
= & \sum_{n, n'} \psi[n] \psi^*[n'] \text{Tr} \left( |n \rangle  \langle n' | \exp \left( - \int_{-\infty}^{+\infty} dk  \xi_M^*(k) \hat{a}(k) \right) \exp \left( \int_{-\infty}^{+\infty} dk \xi_M(k) \hat{a}^\dagger(k)  \right) \right)\nonumber \\
= & \sum_{n, n'} \psi[n] \psi^*[n'] \text{Tr} \left( \prod_k \left[ \hat{a}^\dagger(k) \right]^{n(k)} |0_M \rangle  \langle 0_M | \prod_{k'} \left[ \hat{a}(k') \right]^{n'(k')}  \exp \left( - \int_{-\infty}^{+\infty} dk '' \xi_M^*(k'') \hat{a}(k'') \right) \right. \nonumber \\
& \times \left. \exp \left( \int_{-\infty}^{+\infty} dk'' \xi_M(k'') \hat{a}^\dagger(k'')  \right) \right)\nonumber \\
= & \sum_{n, n'} \psi[n] \psi^*[n'] \text{Tr} \left( |0_M \rangle  \langle 0_M | \prod_{k'} \left[ \hat{a}(k') \right]^{n'(k')}  \exp \left( - \int_{-\infty}^{+\infty} dk '' \xi_M^*(k'') \hat{a}(k'') \right) \right. \nonumber \\
& \times \left. \exp \left( \int_{-\infty}^{+\infty} dk'' \xi_M(k'') \hat{a}^\dagger(k'')  \right) \prod_k \left[ \hat{a}^\dagger(k) \right]^{n(k)} \right)\nonumber \\
= & \sum_{n, n'} \psi[n] \psi^*[n'] \prod_k \left[ \frac{\delta}{\delta \xi_M(k)} \right]^{n(k)} \prod_{k'} \left[- \frac{\delta}{\delta \xi^*_M(k')} \right]^{n'(k')} \text{Tr} \left( |0_M \rangle  \langle 0_M |  \exp \left( - \int_{-\infty}^{+\infty} dk '' \xi_M^*(k'') \hat{a}(k'') \right) \right. \nonumber \\
& \times \left. \exp \left( \int_{-\infty}^{+\infty} dk'' \xi_M(k'') \hat{a}^\dagger(k'')  \right) \right)\nonumber \\
= & \sum_{n, n'} \psi[n] \psi^*[n'] \prod_k \left[ \frac{\delta}{\delta \xi_M(k)} \right]^{n(k)} \prod_{k'} \left[- \frac{\delta}{\delta \xi^*_M(k')} \right]^{n'(k')} \chi^{(-1)}_{0M}[\xi_M, \xi^*_M]\nonumber \\
= & \sum_{n, n'} \psi[n] \psi^*[n'] \prod_k \left[ \frac{\delta}{\delta \xi_M(k)} \right]^{n(k)} \prod_{k'} \left[- \frac{\delta}{\delta \xi^*_M(k')} \right]^{n'(k')} \exp \left( - \int_{-\infty}^{+\infty} dk'' |\xi_M(k'')|^2 \right)\nonumber \\
= & \sum_{n, n'} \psi[n] \psi^*[n'] \prod_k \left[ \frac{\delta}{\delta \xi_M(k)} \right]^{n(k)} \prod_{k'} \left[ \xi_M(k') \right]^{n'(k')} \exp \left( - \int_{-\infty}^{+\infty} dk'' |\xi_M(k'')|^2 \right)\nonumber \\
= & \sum_{n, n'} \psi[n] \psi^*[n'] \sum_{\substack{\mathbf{S} \subseteq \mathbf{K}(n) \\ \mathbf{S}' \subseteq \mathbf{K}(n') }} c(\mathbf{K}(n) \setminus \mathbf{S},\mathbf{K}(n') \setminus \mathbf{S}')  \prod_{k'\in\mathbf{S}'}  \xi_M(k') \prod_{k\in\mathbf{S}} \frac{\delta}{\delta \xi_M(k)}   \exp \left( - \int_{-\infty}^{+\infty} dk'' |\xi_M(k'')|^2 \right)\nonumber \\
= & \sum_{n, n'} \psi[n] \psi^*[n'] \sum_{\substack{\mathbf{S} \subseteq \mathbf{K}(n) \\ \mathbf{S}' \subseteq \mathbf{K}(n') }} c(\mathbf{K}(n) \setminus \mathbf{S},\mathbf{K}(n') \setminus \mathbf{S}')  \prod_{k'\in\mathbf{S}'}  \xi_M(k') \prod_{k\in\mathbf{S}} \left[ - \xi^*_M(k) \right]  \exp \left( - \int_{-\infty}^{+\infty} dk'' |\xi_M(k'')|^2 \right)\nonumber \\
= & \sum_{n, n'} \psi[n] \psi^*[n'] \sum_{\substack{\mathbf{S} \subseteq \mathbf{K}(n) \\ \mathbf{S}' \subseteq \mathbf{K}(n') }} c(\mathbf{K}(n) \setminus \mathbf{S},\mathbf{K}(n') \setminus \mathbf{S}')  \prod_{k'\in\mathbf{S}'}  \xi_M(k') \prod_{k\in\mathbf{S}} \left[ - \xi^*_M(k) \right]  \chi^{(-1)}_{0M}[\xi_M, \xi^*_M].
\end{align}
The general case of Eq. (\ref{A_proof_for_2}) can be proven from the case $p=-1$ and the following identity
\begin{align}
\chi^{(p)}_M[\xi,\xi^*] = & \exp \left( \int_{-\infty}^{+\infty} dK \frac{p+1}{2} |\xi(K)|^2 \right) \chi^{(- 1)}_M[\xi,\xi^*] \nonumber \\
= & \exp \left( \int_{-\infty}^{+\infty} dK \frac{p+1}{2} |\xi(K)|^2 \right) \chi^{(-1)}_{0M}[\xi_M,\xi_M^*] \sum_{n, n'} \psi[n] \psi^*[n'] \nonumber \\
& \sum_{\substack{\mathbf{S} \subseteq \mathbf{K}(n) \\ \mathbf{S}' \subseteq \mathbf{K}(n') }} c(\mathbf{K}(n) \setminus \mathbf{S},\mathbf{K}(n') \setminus \mathbf{S}')  \prod_{k\in\mathbf{S}} \left[ - \xi^*_M(k) \right]  \prod_{k'\in\mathbf{S}'}  \xi_M(k') \nonumber \\
= & \chi^{(p)}_{0M}[\xi_M,\xi_M^*] \sum_{n, n'} \psi[n] \psi^*[n'] \sum_{\substack{\mathbf{S} \subseteq \mathbf{K}(n) \\ \mathbf{S}' \subseteq \mathbf{K}(n') }} c(\mathbf{K}(n) \setminus \mathbf{S},\mathbf{K}(n') \setminus \mathbf{S}')  \prod_{k\in\mathbf{S}} \left[ - \xi^*_M(k) \right]  \prod_{k'\in\mathbf{S}'}  \xi_M(k')
\end{align}

\end{widetext}

\bibliography{bibliography}

\end{document}